\documentclass[12pt,preprint]{aastex}

\slugcomment{Accepted for publication in ApJ}

\begin{document}

\def\ang{{\rm\AA}}

\def\n{\noindent}

\def\ph{\phantom}

\newcommand{\myfigure}[4]{
        \begin{figure*}
        \setbox100=\hbox{
        \epsfxsize=#1 cm
        \epsfbox{#2.ps}}
        \centerline{\hbox{\box100}}
        \caption{#3}
        \label{#4}
        \end{figure*}
}

\title{Detailed Far-UV to Optical Analysis of Four [WR] Stars}

\author{W. L. F. Marcolino\altaffilmark{1}}
\email{wagner@on.br}

\author{D. J. Hillier\altaffilmark{2}}
\email{hillier@pitt.edu}

\author{F. X. de Araujo\altaffilmark{1}}
\email{araujo@on.br}

\and

\author{C. B. Pereira\altaffilmark{1}}
\email{claudio@on.br}

\altaffiltext{1}{Observat\'orio Nacional, Rua Gen. Jos\'e Cristino 77,
  20921-400, \\Rio de Janeiro, Brazil}

\altaffiltext{2}{Department of Physics and Astronomy, University of Pittsburgh, \\
  Pittsburgh, PA 15260 } 

%\bigskip

\begin{abstract}

We present far-UV to optical analyses of four hydrogen deficient 
central stars of planetary nebulae : BD +30 3639, NGC 40, NGC 5315 and 
NGC 6905.  Using the radiative transfer code CMFGEN, 
we determined new physical parameters and chemical abundances for 
these stars. The results were
analyzed in the context of the [WR] $\rightarrow$ PG 1159 evolution via the 
transformed radius-temperature ($R_T \times T_*$) and HR diagrams. We found 
that the use of clumping increases $R_T$ by up to $\sim 0.5$ dex relative
to previous results, whilst our homogeneous models showed systematic
changes of $\sim 0.1-0.2$ dex. NGC 5315 showed itself as an odd object among the previously 
analyzed central stars. Its temperature ($\sim 76kK$) is considerably lower 
than other early-type [WR] stars ($\sim 120-150kK$). From our models for 
NGC 5315 and NGC 6905, it is unclear if early-type [WR] stars 
have smaller C/He mass ratios than other spectral classes, as claimed in the 
literature. In particular, the ratio found for NGC 6905 ($\sim 0.8$) is in rough agreement with 
evolutionary calculations, and with values derived for [WCL] stars. 
We analyzed FUSE spectra of these stars for the first time, and
identified phosphorus in the spectra of BD +30 3639, NGC 40 and NGC 5315 through the 
doublet transition \ion{P}{5} $\lambda\lambda  1118,  1128$ 
($3p\,\,^{2}P^{o}-3s\,\,^{2}S$). The iron, silicon, phosphorus, sulfur and 
neon abundances were analyzed in the context of the nucleosynthesis occurring in previous 
evolutionary phases. We found evidence for an iron deficiency in BD +30 3639 and
NGC 5315, and from fits to the \ion{Si}{4} lines we determined a solar silicon abundance 
for BD +30 3639 and NGC 40. For phosphorus, an oversolar abundance in the
NGC 5315 model was preferred, while in the other stars a solar phosphorus
abundance cannot be discarded. Regarding sulfur, we estimated upper limits 
for its abundance, since no conspicuous lines can be seen in the observed spectra. 
We found that neon is overabundant in BD +30 3639. In the other stars, neon is 
weak or undetectable and upper limits for its abundance were estimated.
Our results are in agreement with theoretical predictions and show the 
usefulness of [WR] stars as testbeds for
nucleosynthesis calculations in the AGB and post-AGB phases.

\end{abstract}


%%%%%%%%%%%%%%%%%%%%%%%%%%%%%%%%%%%%%%%%%%%%%%%%%%%%%%%%%%%%%%%%%%%%%%%%

\section{Introduction}

Hydrogen deficient central stars of planetary nebulae (CSPN) present 
a challenge to stellar evolution theory: it is still unclear how a star 
removes its hydrogen when going from the AGB to the planetary 
nebula stage. One of the first theories trying to explain this phenomenon was 
proposed by Iben et al. (1983). In their {\it born-again} scenario, a very late thermal 
pulse (VLTP) occurs in the white dwarf cooling track of the HR diagram.
As a consequence, all the remaining hydrogen is burned and the star is driven 
to red giant dimensions, near the AGB, for a second time. Two other thermal pulses could
also potentially give rise to hydrogen deficient CSPN. One is the AFTP (AGB final thermal pulse), 
occurring at the end of the AGB phase, and the other is the LTP (late thermal
pulse), which occurs when the star is already on the post-AGB track. It seems
that all three scenarios are necessary in order to explain the wide range of hydrogen 
deficient CSPN properties (e.g., hydrogen abundance and nebular diameters), and
that there is no unique mechanism or single channel of evolution
(Bl\"ocker 2001; Herwig 2001). 

Binarity could also provide an avenue for the creation of hydrogen deficient CSPN. 
De Marco \& Soker (2002), for example, raised the interesting possibility of a 
companion engulfed by an AGB star. In this way, extra mixing would be induced 
and a hydrogen deficiency would be achieved. However, the scenario is still 
speculative and lacks quantitative calculations. 

The evolutionary sequence of hydrogen deficient CSPN is also uncertain. Generally
these objects are divided into three classes : [WR], [WC]-PG 1159 and PG 1159 stars. [WR] 
stars show spectra very similar to Wolf-Rayet stars of Pop. I, with strong and 
broad emission lines. [WC]-PG 1159 stars show both 
absorption and weak emission lines in their optical spectra, while wind 
features are seen in the UV. Parthasarathy et al. (1998) claim that
[WC]-PG 1159 and the {\it weak emission line stars} ([WELS]), presented in the work of Tylenda
et al. (1993), constitute the same class. PG 1159 stars show mainly absorption 
lines in their spectra.

Analysis of the spectra of the central stars or of their nebulae seems to 
indicate the following evolution : [WCL] $\rightarrow$ [WCE] $\rightarrow$
[WC]-PG 1159 = [WELS] $\rightarrow$ PG 1159 (Werner \& Heber 1991a,b; Zijlstra 
et al. 1994; Gorny \& Tylenda 2000; Pe\~na et al. 2001; Leuenhagen et al. 1996; 
Koesterke \& Hamann 1997a,b; Leuenhagen \& Hamann 1998). However, there are still fundamental 
questions to be addressed. A major problem is the C/He mass ratio obtained for early-type [WR]
stars, which is $\sim 0.35$ on average. This is lower than found in the other classes, and
is thus in contradiction to the above evolutionary sequence. It also disagrees
with evolutionary models which find this ratio to be $\sim 1$ (Herwig 2001). A
second problem comes from the work of Gorny \& Tylenda (2000), who showed that some
stars might not follow this evolution channel. Through an analysis of nebular parameters, the 
authors suggest that some objects that have experienced the VLTP could 
evolve directly to the PG 1159 stage without becoming [WC] stars. Another 
important problem is that the role of the [WELS] is not yet understood 
(Marcolino \& de Ara\'ujo 2003). Indeed,  Pe\~na et al. (2003a;b) were 
able to show that nebular expansion velocities of [WC] stars are generally 
higher than for the [WELS], in contradiction to that expected from
[WC] $\rightarrow$ [WELS] evolution.
Moreover, the presence of hydrogen is still questionable (De Marco \& Barlow
2001) and can favor a specific thermal pulse model. While in the AFTP and LTP 
a small amount is found, in VLTP models the hydrogen is completely burnt.

So far, the majority of [WR] stars analyzed by means of non LTE expanding 
atmosphere models is due to the Potsdam group code (see Koesterke 2001 and 
references therein). Additional studies of three [WCL] stars were undertaken by
De Marco \& Crowther (1998, 1999) using an independent code (CMFGEN, 
Hillier \& Miller 1998). All objects were studied without metal line-blanketing or
clumping (exceptions are the recent work of Stasi\'nska et al. 2004 and of Pe\~na et
al. 2004 regarding two stars on LMC, Herald \& Bianchi 2004a,b, which included 
one [WC4] and two [WC]-PG 1159 stars and Crowther et al. 2006, who
compared the [WC9] star BD +30 3639 with the Wolf-Rayet star HD 164270). 
These two effects have been shown to be very important when analyzing 
hot stars with winds (Hillier \& Miller 1998, 1999; Gr\"afener et al. 2002), 
having an impact on important physical parameters (e.g., $T_{eff}$ and
$\dot{M}$). 

With the aim of addressing some of the problems and uncertainties
discussed above, we have derived physical parameters and chemical abundances of 
a small sample of H deficient CSPN (BD +30 3639, NGC 40, NGC 5315 and NGC 6905) 
using the CMFGEN code developed by Hillier \& Miller (1998). Contrary to most 
of the previous works on CSPN, line-blanketing and clumping are taken into 
account in the models. We also rely on improved atomic data and fits to 
larger spectral intervals, including the FUSE region now available for 
some CSPN. Among the issues that we address are : the influence of
line-blanketing and clumping on the evolutionary sequence; the peculiar 
values of the C/He mass ratio for the early-type [WR] stars and the iron 
abundance and its possible depletion as predicted by AGB and post-AGB 
evolutionary calculations. 

This paper is organized as follows. The observational data is described in 
Section 2. In Section 3 we explain the code used (CMFGEN) and its main 
assumptions, while in Section 4 we present our analysis from the far-UV to 
the optical for each object. In Section 5 we discuss our results. 
Finally, in Section 6 we summarize the main points of our work.

%%%%%%%%%%%%%%%%%%%%%%%%%%%%%%%%%%%%%%%%%%%%%%%%%%%%%%%%%%%%%%%%%%%%%

\section{Observations}

The optical data of BD +30 3639 and NGC 5315 were obtained in 2000 
using the Boller \& Chivens spectrograph at the Cassegrain focus on the 1.52m 
ESO telescope in La Silla, Chile. The resolution is about 2\AA\, in the
spectral range $\sim 4000-6000$\AA. The spectra were reduced using IRAF.
Default procedures such as bias and sky subtraction, spectral extraction and 
flat-field corrections were taken into account. He, Ar and Fe lamps were used
to provide wavelength calibration for every stellar frame. More details can be 
found in our previous paper (de Ara\'ujo et al. 2002). The spectra of NGC 40 and NGC 6905 were
kindly provided by Paul A. Crowther. They were observed on the 2.5m Isaac
Newton Telescope (INT) in La Palma, Spain. The spectral interval is $\sim
3600-6800$\AA\, and the resolution is $\sim 1.5$\AA\, for NGC 40 and $\sim 3$\AA\,
for NGC 6905. Further details can be found in Crowther et al. (1998).

In the ultraviolet and far-ultraviolet region, we used public data from the
Multi-mission Archive at STScI (MAST) \footnote{http://archive.stsci.edu/} from
the IUE and FUSE satellites. The wavelength coverage is $1087-1182$\AA\, 
(also $980-1182$\AA\, in NGC 6905) for FUSE spectra and $\sim 1100-3200$\AA\, for 
IUE spectra. All FUSE spectra were co-added and smoothed with a 3-5 point
average for a better signal-noise ratio and clarity, respectively. 
The resolution of the models shown in the next sections was chosen to 
match the resolution in each observed spectrum.

%%%%%%%%%%%%%%%%%%%%%%%%%%%%%%%%%%%%%%%%%%%%%%%%%%%%%%%%%%%%%%%%%%%%%

\section{Model atmosphere}

The physics present in hydrogen deficient CSPN is very complex. 
CSPN have high effective temperatures ($\sim 20-200kK$), intense radiation 
fields and expanding atmospheres. These necessitate, at a minimum,
the use of spherical expanding non LTE model atmospheres. 
Moreover, the presence of clumps (enhanced density regions, i.e., {\it clumping}) is 
predicted by radiation hydrodynamic theory (Owocki 1994) and supported by 
observational studies (see for example Bouret et al. 2005 and references
therein) which invalidates the commonly adopted hypothesis of a smooth outflow.

For our analyses we used the non LTE radiative transfer code, CMFGEN (Hillier \&
Miller 1998). CMFGEN solves the radiative transfer equation, in a spherically symmetric expanding
outflow, simultaneously with the statistical and radiative equilibrium equations. 
It has been applied to the study of several classes of objects where non LTE
conditions and a stellar wind are present (e.g., WR, LBV and O stars). More
recently, CMFGEN was used to investigate the photospheric phase of type II 
supernovae (Dessart \& Hillier 2005).

Line-blanketing is known to change the atmospheric structure of WC stars 
and it affects important lines such as \ion{C}{3} $\lambda 5696$ and \ion{C}{4} $\lambda \lambda
5801, 5811$, which are used in classification schemes (Crowther et al. 1998;
Acker \& Neiner 2003). To facilitate the inclusion of metal 
line-blanketing in CMFGEN, {\it super-levels} are used. In this formalism, levels with similar properties 
are treated as one and have the same departure coefficient, saving a
considerable amount of computer memory and time. In the atomic model for
\ion{Fe}{4}, for example, 1000 levels could be reduced to 100. Models with 
different super-levels assignments were tested in order to check for differences in 
the theoretical spectra. 

Clumping is incorporated into CMFGEN using a volume filling factor ($f$) approach (Hillier \& Miller 1999).
In all models of this work we assume $f=0.1$. Evidence for clumping in spectral fits can be primarily seen in the 
red wings of emission lines, although a few individual line strengths and profiles 
can also show some sensitivity. Clumping tends to reduce the derived mass-loss rates
by a factor of $\sim 3-5$. While a non-clumped wind can be ruled out by the
modeling, it is difficult to derive an accurate clumping value. Consequently, 
we always tabulate both the derived mass-loss rate and the unclumped value 
$\dot{M}_{uncl}$ (=$\dot{M}/\sqrt{f}$). Models with fixed $\dot{M}_{uncl}$ 
generally show very similar spectra. In some objects of our sample (BD +30 3639 
and NGC 40), direct evidence for clumping has been obtained from observations 
of moving features on the top of some specific lines (see for example, 
Grosdidier et al. 2000). 

In order to improve the modeling of \ion{C}{2} lines we implemented new
autoionization data made available by the Auburn-Rollins-Strathclyde 
collaboration\footnote{Auburn-Rollins-Strathclyde Dielectronic Recombination 
Data - http://www-cfadc.phy.ornl.gov/}, which is not included in Opacity
Project Calculations.

\section{Analysis}

The main input parameters in a CMFGEN model are the luminosity ($L$), 
the radius ($R_*$), the mass-loss ($\dot{M}$), the velocity law 
[$v(r)$] and the chemical abundance. The stellar temperature $T_*$ follows 
from the relation $L=4\pi R_* ^2 \sigma T_*^4$ and the effective temperature $T_{eff}$ 
is defined where the Rosseland optical depth is 2/3. 
In order to explore these parameters in a efficient way, we adopted the 
following strategy for our analyses. As the distances involved are not 
precisely known, we chose to fix the luminosity in $5000L_{\sun}$ in our 
models. This particular choice of $L$ was made as it is the approximate 
value found from the evolution of a main sequence star of $\sim 3M_{\odot}$ 
(Bl\"ocker 2001; Althaus et al. 2005). With the luminosity fixed, the 
distance was derived from the continuum spectral fit. 
If a reliable distance is determined for one of the stars of our sample,
the physical parameters derived by our analysis can be scaled to new 
values by keeping the so-called {\it transformed radius} and $T_{eff}$ 
constant (Schmutz et al. 1989; Hillier \& Miller 1999; see
\S\ref{sect_ev_seq}). We then investigated different stellar 
temperatures by changing the radius of the star. The velocity law used was a simple 
$\beta$-law ($\beta=1$) and $v_{\infty}$ was fixed by the blue 
shifted absorption in a P-Cygni profile or by values found in 
the literature. Thereafter, we changed the mass-loss until we 
obtained a satisfactory fit. Abundance changes were the last step on the 
modeling process. The C/He ratio was derived mainly from the fit to the \ion{He}{2} $\lambda
5412$ and \ion{C}{4} $\lambda 5471$ transitions. In some cases we found that
the best fit to these lines yielded a poor fit to other lines
(e.g. \ion{C}{3} $\lambda 5696$). In such cases
we chose to determine the abundance by the overall spectral fit.
Because a perfect match to the observed spectra is not possible, 
a quality judgment was exercised to choose our final models. 

We generally started with simple atomic models (e.g., He, C, O and Fe ions) and we 
gradually added and tested the influence of new elements on the spectrum. 
This approach helped to determine and fit previously unidentified 
features, such as \ion{P}{5} and \ion{Al}{3} in the ultraviolet. 
We did not include hydrogen in our analysis. Its presence in [WR] stars is very 
difficult to determine and it is still a matter of debate.
The ions considered in the final model for BD +30 3639 are shown in Table 1 
as an example. We included a total of 4428 full-levels, which were reduced 
to 1391 with the use of super-levels. For the other stars, only a few 
modifications were done to the atomic model. We now discuss the analysis of the
individual stars: BD +30 3639 (\S 4.1), NGC 40 (\S 4.2), NGC 5315 (\S 4.3) and
NGC 6905 (\S 4.4).

%%%%%%%%%%%%%%%%%%%%%%%%%%%%%%%%%%%%%%%%%%%%%%%%%%%%%%%%%%%%%%%%%%%%%

\subsection{BD +30 3639}

BD +30 3639 (also called Campbell's star) is a well studied
planetary nebula that has a [WC9] central star. It is one of the few planetary 
nebulae with detected X-ray emission (Arnaud et al. 1996; 
Kastner et al. 2000; Maness et al. 2003), and, unusually, it has both oxygen and 
carbon-rich dust envelopes (Waters et al. 1998).  Further, it has
high velocity molecular knots: the first detected in a planetary nebula (Bachiller et al. 2000).
Despite efforts to understand its nature and evolution only a few studies have 
focused on its central star.

Leuenhagen et al. (1996) were the first to use non LTE expanding 
atmosphere models to obtain, based primarily on optical line fits, 
physical parameters and chemical abundances for BD +30 3639. 
Their model atmosphere considered only H, He, C and O ions, and 
at that time, they were unable to include line-blanketing and clumping. 
The presence of clumping in BD +30 3639 was reported by Acker et al. (1997)
and Grosdidier et al. (2000) through the study of moving features on the top of the 
\ion{C}{3} $\lambda 5696$ line. 

Recently, Crowther et al. (2006) also analyzed this star using CMFGEN.
However the study primarily focused on the Population I Wolf-Rayet star HD 164270, 
and a comparison was made with BD +30 3639 because of their identical
spectral classification (WC9). Here emphasis is given to the determination 
of physical parameters, chemical abundances and the star's evolutionary state.

\subsubsection{Optical and ultraviolet fit}

In Fig. 1 we show the observed and theoretical optical spectra. 
The distance derived from our fit is 1.2kpc, which is the same value 
found by Li et al. (2002) using Hubble Space Telescope observations of 
the expansion of the nebula of BD +30 3639. In the $4000-5000$\AA\, range 
we have a reasonable agreement between model and observations, but we had
difficulties in fitting the observed $4650$\AA\, profile simultaneously with the other
features. This blend is formed by \ion{C}{3} and \ion{C}{4} transitions and it is weaker 
in the model than in the observed spectrum. On the other side, our model provides a very good
fit to the $5000-6000$\AA\, interval. The \ion{C}{3} $\lambda
5696$ line was only reproduced after the inclusion of clumping (with $f =
0.1$). The \ion{He}{1} $\lambda 5876$ profile is contaminated by the nebula.
In this case, a resolution higher than 2\AA\ would be necessary in order 
to better separate wind and nebular emissions. 

The far-ultraviolet spectrum from FUSE and our model are shown in Fig. 2
(top). We used only the LiF2A channel ($\sim 1087-1182$\AA) which 
is free of severe interstellar contamination. This is the first spectral analysis of 
this object in this region. Our main findings are the P-Cygni profiles in $\lambda 1118$ and $\lambda
1128$, which were only fitted after the inclusion of phosphorus (\ion{P}{5}).
The determination of the phosphorus abundance is important in the context 
of nucleosynthesis during the AGB evolution and will be discussed later in the
paper. Our models also indicate a silicon contribution in 1128\AA\, from \ion{Si}{4}.

The high-resolution spectrum from the IUE satellite is also shown in Fig. 2 
along with our model. The observed continuum near $1200$\AA\, does not match 
between IUE and FUSE spectra. The FUSE continuum is higher $\sim5 \times
10^{-13}$ ergs cm$^{-2}$ s$^{-1}$ \AA$^{-1}$.                                          %%%%% see models/bd30/iuefuse*.ps 
We could only achieve agreement between observed and      
theoretical spectrum from the far-UV to the optical using slightly 
different values for E(B-V) ($\sim 0.34-0.40$). Errors in flux calibrations 
and uncertainties in the extinction law are probably reasons for the
discrepancy. It is known, for example, that BD +30 3639 has a dusty halo 
which contributes to its own extinction (Harrington et al. 1997). 

The most prominent features in the $\sim 1100-2000$\AA\, interval are :
\ion{Si}{4} ($\lambda 1394$, $\lambda 1403$), \ion{C}{2} ($\lambda 1335$) and 
\ion{C}{3} ($\lambda 1246$, $\lambda 1308$, $\lambda 1909$ and $\lambda 1923$). 
We highlight the presence of several lines between $\sim 1400-1900$\AA. According 
to our models they stem mainly from \ion{Fe}{4} transitions. Indeed, if we ignore 
iron, a flat spectrum is produced in this region. Thanks to the high
resolution used, the spectrum after $\sim 1850$\AA\, could be matched after 
the inclusion of \ion{Al}{3}, which gave origin to the transitions $\lambda
1855$ and $\lambda 1863$ ($3p\,\,^{2}P^{o}-3s\,\,^{2}S$). 

In Fig. 3 we show our model and the high-resolution IUE spectrum in the 
$\sim 2000-3000$\AA\, range. Unfortunately, the signal-noise ratio of this 
region is not so good, although several lines could be identified : 
\ion{C}{3} (e.g., $\lambda 2010$, $\lambda 2092$, $\lambda 2163$, 
$\lambda 2296$ and $\lambda 2726$), \ion{C}{4} (e.g., $\lambda 2405$, 
$\lambda 2524$, $\lambda 2530$, $\lambda 2698$), \ion{Ne}{3} ($\lambda 2553$
and $\lambda 2678$), and two \ion{He}{2} lines 
($\lambda 2511$ and $\lambda 2733$). The \ion{S}{5} $\lambda 2654$ line 
is indicated in the plot and will be discussed together with the neon lines 
later in the paper. The $2950-3100$\AA\, interval presents 
important features and it was used to derive the oxygen abundance.

%%%%%%%%%%%%% NEBULAR LINES 
% optical : Hbeta, Hgamma, Hdelta, [OIII] 4959, [OIII] 5007,
% [N I] 5199, [Cl III] 5538, [N II] 5754
% between 1200-2000 : none
% between 2000-3000 :
%    *  after C III 2296 we see several narrow intense lines... they are nebular :
%    *  C II] 2323.5, 2324.7, 2325.4, 2326.9, 2328.1
%    *  [OII] 2470.96

\subsubsection{Physical parameters and chemical abundances}
% final model : af17

The basic parameters of our final model are shown in Table 2. 
In addition, we also show in Table 3 the wind efficiency $\eta$ ($=\dot{M}v_\infty c/L$) and 
the blanketed ionizing flux shortward of $912$\AA\, (Q$_o$), $504$\AA\, (Q$_1$) and
$228$\AA\, (Q$_2$). These values can be useful in future nebular investigations.

There are some differences between our results for BD +30 3639 and the work by Leuenhagen et
al. (1996).  Their parameters are also in Table 2, and are scaled for the
distance of 1.2kpc. Regarding the mass-loss, our value is $5 \times 10^{-7}
M_{\sun} / {\it yr}$, almost one order of magnitude less than their value,
which is $4\times 10^{-6}M_{\sun}/ \it{yr}$. Even without clumping our result 
is significantly lower, $ \dot{M}/\sqrt{f} = 1.6 \times 10^{-6}M_{\sun}/ \it{yr}$. 
It is important to note that their definitions of star temperature ($T_{*}$)
and effective temperature ($T_{eff}$) are different from ours. 
Our $T_{eff}$ is the temperature where the Rosseland optical depth is $2/3$.
This temperature is called $T_{2/3}$ in their work and they found $42000K$. 
Our $T_{*}$ is defined where the Rosseland optical depth is $20$. They call 
it the effective temperature and found $47000K$. The reason for quoting
two temperatures is that $T_{2/3}$ is often strongly affected by the wind
properties.

At first glance, our results are somewhat different from those recently published 
by Crowther et al. (2006), who also used the CMFGEN code and the same 
distance (d = 1.2kpc) derived by us. It must be kept in mind however that 
we have adopted throughout this work a fixed value for the luminosity 
($L =  5000L_\odot$, on the basis of evolutionary results already cited) while 
their model for BD +30 3639 has $L \sim 6000 L_\odot$. Our derived $T_*$ differs 
by about 15\% from theirs and our $T_{eff}$ differs by only about 3\%. 
Moreover, their physical parameters have been derived basically from the
fitting of the lines \ion{C}{2} $\lambda 4267$, \ion{C}{3} $\lambda 
5696$ and \ion{C}{4} $\lambda \lambda 5801, 5812$, while we have chosen to 
utilize the whole spectrum. Another reason for the differences found is the 
slightly different visual magnitude obtained by us, which is V$ = -1.3$, as well 
as a different extinction correction. Given these uncertainties, we 
conclude that the results shown in Table 2 are not incompatible. 

The chemical abundance of the final model is $\beta_{He} = 43\%$, $\beta_{C} = 51\%$ and
$\beta_{O} = 6\%$ (by mass), which corresponds to C/He = 0.4 and O/He = 0.035
(by number). In order to test the sensitivity of our models to abundance changes we 
ran models with different values for C/He, fixing O/He at 0.035.
A reasonable range for the C/He ratio in this object is 0.2 $\leq$ C/He $\leq$ 0.6. 
The \ion{O}{3} lines in the $2950-3100$\AA\, interval (see Fig. 3) were used 
to determine the O/He ratio. In order to check the presence of nitrogen, 
\ion{N}{2}\ion{-}{4} was added during test models. We could not identify 
any conspicuous transitions in the observed spectra. Using lines such as 
\ion{N}{3} $\lambda 4097$ we can conclude that its abundance is $\beta_N < 0.2\%$. 

% 
% uncertainty in the abundance determination
%
% af22    He = 55 / C = 36 / O = 8    (roughly 30% for each element)
%
% af23    He = 34 / C = 61 / O = 5
%
% af17    He = 43 / C = 51 / O = 6    (our best model)
%

We were able to estimate a range for the iron abundance of 
$0.13 \lesssim \beta_{Fe}/\beta_{Fe_{\odot}} < 1$. 
In Fig. 4 (top) we can see that a solar value results in stronger lines 
compared to the observations. On the other side, a value of 6\% of 
the solar abundance (bottom) is too low. Our conclusion is that BD +30 3639 shows a 
deficiency in iron, but the exact value is difficult to determine. 
Our best fit uses 25\% of the solar iron abundance.
Similar results were also found in other works (see Miksa et
al. 2002 and references therein). These facts support theoretical
calculations which predict that iron is depleted by neutron capture 
occurring during and after the AGB phase (see Herwig et al. 2003; Werner \&
Herwig 2006). According to Werner \& Herwig (2006), some
of the iron is converted into 
nickel and a definite signature of its depletion would be the reduction of 
the Fe/Ni ratio below the solar value ($\sim 20$), to $\sim 3$. We tested models 
including Ni III-V, but we were unable to determine the nickel abundance precisely. 
Therefore, the most we can say is that our result favors the expectation that 
iron is depleted, but a quantitative comparison involving nickel is not 
possible at present. 

% nitrogen : models af5, af5b, nit (based on a41c)

% iron *********************
% top : Fe = Fe(sun)       (model a41c)
% 2nd : Fe = 0.25  Fe(sun) (model af9b)
% 3rd : Fe = 0.060 Fe(sun) (model af9d)
%
% another test : Fe = 0.125 Fe(sun) (model af9c) (not in the plot)

% nickel models :
% nickel/ and nickel2/

% Fig 5 CAPTION :
% top    : model a41c  5dec96 OIII data, O/He=0.1,    50 super-levels
% middle : model af7b 20jun01 OIII data, O/He=0.1,   267 super-levels
% bottom : model af7c 20jun01 OIII data, O/He=0.035, 267 super-levels

% models to check uncertainty in oxygen - these limits are reliable
% af20 : O/He = 0.049
% af21 : O/He = 0.021
% af20b: O/He = 0.070 (upper value)
% af21b: O/He = 0.010 (lower value) 
% all based on af17 
%
% O/He = (0.04 +- 0.03) is reliable estimate

%%%%%%%%%%%%%%%%%%%%%%%%%%%%%%%%%%%%%%%%%%%%%%%%%%%%%%%%%%%%%%%%%%%%%

\subsection{NGC 40}

The central star of NGC 40 is classified as [WC8] (Crowther et al. 1998). 
Its effective temperature is quite controversial. On one hand, Bianchi \&
Grewing (1987) claimed a value of $\sim 90000K$ from the analysis of
UV spectra. With the use of a non LTE expanding atmosphere code, 
Leuenhagen et al. (1996) derived $T_* \sim 78000K$ and $T_{eff} \sim 46000K$.
In contrast to these results, the nebula presents a low excitation class 
compatible with a central star temperature of only $\sim 38000K$ 
(Pottasch et al. 2003). Thus, it seems that the nebula does not perceive 
the central star temperature. Based on these facts, Bianchi \& Grewing (1987) 
proposed the existence of a carbon curtain screening high energy photons.
A qualitative UV analysis was also presented by Feibelman (1999), where a terminal velocity 
of $\sim 1700$ km $s^{-1}$ was derived from the \ion{C}{4} $\lambda 1549$ line. 
This study also reinforced a high temperature of $\sim 90000K$ by fitting the continuum with a
blackbody curve and based on the presence of the stellar \ion{He}{2} $\lambda 1640$. 

\subsubsection{Optical and ultraviolet fit}

In Fig. 5 we show our model and the observed optical spectrum. 
The value derived for the distance is 1.4 kpc. For the interstellar 
reddening we use an E(B-V) of 0.41. The \ion{C}{3} $\lambda 4650$
line is the most intense in the entire spectrum. 
Other important features are \ion{He}{2} $\lambda 4686$, \ion{He}{2} $\lambda 5412$, 
\ion{C}{4} $\lambda 5471$, \ion{C}{3} $\lambda 5696$ and \ion{C}{4} $\lambda\lambda 5801, 12$. 
In contrast to BD +30 3639, where \ion{C}{3} dominates,
\ion{C}{4} starts to compare in strength with \ion{C}{3} and \ion{C}{2} is weak 
(e.g., $\lambda 4267$) or absent.  This suggests that NGC 40 
has $T_{eff}$ higher than 48000K (BD +30 3639's value), in contrast 
with low values found from nebular techniques (Pottasch et al. 2003).
We will see quantitative details in the next section.

Fortunately, FUSE spectra are also available for NGC 40. No quantitative study 
of the FUSE spectral region has been previously done for this star. 
The spectrum in the interval 
$1087-1182$\AA\, (LiF2A channel) is shown in Fig. 6 (top) along with our model. 
Despite having a higher temperature than BD +30 3639, 
we still had to include \ion{P}{5} to fit the features in $\lambda 1118$ and $\lambda
1128$. Other lines identified with the help of our models are \ion{C}{3} $\lambda
1175$, \ion{C}{4} $\lambda 1108$ (which might have interstellar absorption lines 
contaminating it) and \ion{C}{4} $\lambda 1169$. 

The fit to the IUE spectrum is also presented in Fig. 6. High-resolution data 
was used in the $\sim 1000-2000$\AA\, interval. Some conspicuous features are
: \ion{He}{2} $\lambda 1640$, \ion{Si}{4} $\lambda 1722$, \ion{C}{3} $\lambda
1909$ and \ion{C}{3} $\lambda 1923$. The \ion{Si}{4} $\lambda 1722$ and
\ion{S}{5} $\lambda 1502$ lines will be discussed in Section 5.

We did not get a very good fit to \ion{C}{4} $\lambda 1549$ even 
using a Voigt profile. Difficulties in the fitting of this feature 
have already been reported before and it is known that this line is very sensitive 
to clumping (Hillier \& Miller 1999; Crowther et al. 2002). However, a decrease of 
the filling factor $f$ from 0.1 to 0.03 improves the fit, but only regarding 
its intensity. A very low value of $0.01$ was tried, but is not so different 
from the $0.03$ fit and it is probably too extreme. With a
lower mass-loss we can get a better profile, but we have a drawback 
regarding the fit quality in other parts of the spectrum. The 
reduced mass-loss improves the fit since it reduces the optical depth of 
\ion{C}{4} $\lambda 1549$. The optical depth in this line is large, 
and as a consequence the Voigt wings are important for the formation of 
the observed profiles. The neglect of partial redistribution effects, 
which might be significant in such a case, is likely to contribute to the discrepancy.

A low resolution IUE spectrum was used in the $\sim 2000-3000$\AA\, range 
(bottom of Fig. 6). The \ion{C}{3} $\lambda 2296$ theoretical line is too 
strong. Unfortunately, changes of the model parameters do not lead to better 
agreement without adversely affecting the fit to other features. 
The remaining transitions present a very reasonable fit.

\subsubsection{Physical parameters and chemical abundances}
% ``THE'' final model is n32

The parameters for NGC 40 are shown in Table 2. Scaling the results of
Leuenhagen et al. (1996) to our distance, we find : $R_{*} = 0.46 R_{\sun}$, 
$L = 7450 L_{\sun}$ and $\dot{M} = 4 \times 10^{-6} M_{\sun} / {\it yr}$. 
Their higher luminosity is reflected in their higher value for $T_*$. 
As for BD +30 3639, our derived mass loss rate ($\dot{M}/\sqrt{f}$) 
is approximately a factor of 2 lower than that found using the
Potsdam model.

As mentioned earlier, the temperature of NGC 40 is controversial.
On the one hand, nebular techniques such as the Zanstra and Stoy method provide 
a temperature around $38000K$ (Pottasch et al. 2003). On the other hand,
analysis of the central star seems to indicate temperatures up to $\sim 90000K$.
In order to address this issue, we ran several models with different 
effective temperatures. From the response of optical and UV lines 
(e.g., \ion{C}{4} $\lambda 1549$, \ion{C}{3} $\lambda 4650$, \ion{He}{2}
$\lambda 4686$, \ion{C}{3} $\lambda 5696$ and \ion{C}{4} $\lambda \lambda
5801, 12$) to temperature changes, we conclude that NGC 40 has $T_{eff} = (71
\pm 10) kK$. A possible reason for the low central star temperatures found by 
nebular techniques is the neglect of a stellar wind and the use of the 
black-body approximation. The number of ionizing photons emerging to 
the planetary nebula can deviate significantly from that of the inner parts 
of the stellar wind. This can result in a planetary nebula perceiving the 
central star as a low temperature object. For example, for a star with 
$T_* = 38000K$ which radiates like a black-body (BB), the number of ionizing 
photons per second are Log $Q^{BB}_0 = 47.37$, Log $Q^{BB}_1 = 46.33$ and Log $Q^{BB}_2 =
43$. Our results, for $T_* \sim 73kK$ are : Log $Q_0 = 47.56$, Log $Q_1 = 46.91$ and Log $Q_2 =
36.79$. As can be seen, our values for \ion{H}{1} and \ion{He}{1} are 
compatible with the BB values and for \ion{He}{2} our model indicates a much
lower number of photons. Therefore, the number of ionizing photons escaping 
from the wind is consistent with (or even less than) the number of ionizing 
photons of a star with a much lower temperature, if it radiates like a 
black-body (as assumed in some nebular methods). A consistent analysis of the 
wind and the nebula, with the stellar parameters obtained here as input
parameters on a photoionization model will certainly help clarify this problem.

%
% Teff analysis can be found on n49 (Teff=77480K) and n50 (Teff=58140K)
%
%
% C/He models :
%
% n38 : C/He = 0.24     O/He = 0.035 ***** good for comparison
% n38b: C/He = 0.15     O/He = 0.035
% n37 : C/He = 0.56     O/He = 0.035
% n37b: C/He = 0.68     O/He = 0.035 ***** good for comparison
% n37c: C/He = 0.84     O/He = 0.035
% n37d  C/He = 0.97     O/He = 0.035
% models based on n32
%
% O/He models : (reasonable limits for oxygen)
% 
% n47 : O/He = 0.07      C/He = 0.4
% n48 : O/He = 0.01      C/He = 0.4
%
% models based on n32

The chemical abundance of the final model is $\beta_{He} = 43\%$, $\beta_{C} = 51\%$ and 
$\beta_{O} = 6\%$ (by mass), which corresponds to C/He = 0.40 and O/He = 0.035
(by number). Decreasing C/He below this value improves the fits to \ion{He}{2}
$\lambda 5412$ and \ion{C}{4} $\lambda 5471$, but \ion{C}{3} $\lambda 5696$ starts to get
too weak. Our uncertainty estimate for C/He is the same as for BD +30 3639,
i.e., 0.2 $\leq$ C/He $\leq$ 0.6. We used \ion{O}{3} $\lambda 5593$ and the \ion{O}{3} lines in 
the $2950-3100$\AA\, interval as the diagnostic lines for the determination of
the O/He ratio. Nitrogen features are not observed in this object. If
present, its abundance in mass fraction is $\beta_N < 0.1\%$ and it does not 
influence the spectrum. By including \ion{N}{3}$-${\small\rmfamily V} 
we could verify for example, 
that \ion{N}{4} $\lambda 1718$ appeared and it does not exist in the IUE spectrum.
Regarding iron, it is difficult to say if this object has an abundance lower 
than solar. An ultraviolet synthetic spectrum in the $1000-2000$\AA\, region
with a solar iron abundance is not different enough from a lower value 
($\sim 80\%$ less) to allow a firm conclusion. 

%%%%%%%%%%%%%%%%%%%%%%%%%%%%%%%%%%%%%%%%%%%%%%%%%%%%%%%%%%%%%%%%%%%%%

\subsection{NGC 5315}

NGC 5315 is classified as a [WC4] star by Crowther et al. (1998) and as a
[WO4] star by Acker \& Neiner (2003). The planetary nebula is compact 
($\sim 4\arcsec$), almost spherical and was studied 
from the infrared to the ultraviolet (see for example, Pottasch et al. 2002 and
Peimbert et al. 2004). So far, no spectroscopic analysis of the
central star has been performed using a non LTE code.  Feibelman (1998) presented 
a qualitative analysis based on IUE data and obtained a terminal velocity of $3600$ km $s^{-1}$ and 
$T_{*} \sim 80000K$. de Freitas Pacheco et al. (1986; 1993) derived $T_{*} \sim 82700K$ and a 
terminal velocity of $2600$ km $s^{-1}$. However, Pottasch et al. (2002)
claim a lower temperature of $\sim 66000K$ based on the Zanstra method for 
hydrogen and on the absence of \ion{He}{2} lines in the nebula. 
As we mentioned previously, nebular methods that assume black-body 
radiation from [WR] stars should not be used to derive their temperatures.

\subsubsection{Optical and ultraviolet fit}

Early-type [WR] stars usually have intense mass-losses and high terminal velocities 
($v_{\infty} \sim 1000-3500$ km $s^{-1}$). The latter complicates their modeling 
because many spectral features are severely blended. We highlight in particular 
the $\lambda 4650$ and $\lambda 5805$ features seen in Fig. 7, where we
compare the synthetic and observed spectrum. According to our model, 
the feature at $\lambda 4650$ is composed of \ion{He}{2}, \ion{C}{3} 
and \ion{C}{4}, whilst $\lambda 5805$ is pure \ion{C}{4}. The distance 
used is 2.5kpc and the interstellar reddening is E(B-V) = 0.37 (Pottasch et
al. 2002). 

The far-ultraviolet fit to the FUSE spectrum is shown in Fig. 8 (top).
Again, we highlight the presence of the \ion{P}{5} lines  $\lambda 1118$ 
and $\lambda 1128$. The fit presented uses four times the solar phosphorus 
abundance and it will be discussed further in Section 5. The transitions 
\ion{C}{3} $\lambda 1175$ and \ion{C}{4} $\lambda 1108$ and $\lambda 1169$
could also be identified. 

For the UV analysis we used low resolution IUE spectra. Our fit is shown in 
Fig. 8. The most intense lines are indicated in the plot. 
The observed spectrum in the $\sim 2000-3000$\AA\, interval was obtained
in the IUE small aperture mode and it was scaled by a factor of 2.5. 
We decided not to use the other available large aperture spectra because they 
seem to show nebular continuum contamination beyond 2500\AA\,. 
The compactness of the planetary nebula and the large aperture 
of $10''\times 20''$ support this idea. Unfortunately, the signal-noise ratio 
in $\sim 2000-3000$\AA\, is not so good. Nevertheless, our model allowed us 
to identify the presence of \ion{C}{4} $\lambda 2405$, \ion{C}{4} $\lambda 2530$ and 
\ion{C}{3} $\lambda 2296$. Feibelman (1998) considered this last feature 
as having a nebular origin, but our fit suggests otherwise.

\subsubsection{Physical parameters and chemical abundances}
% best model : /phosp

The parameters of our final model for NGC 5315 are shown in Table 2. 
If compared to de Freitas Pacheco et al. (1986; 1993), our unclumped mass-loss is 
virtually the same. However, their radius and luminosity ($L = 3924 L_{\sun}$)
are different, implying a higher $T_*$. 
It is interesting to note that the majority of the early-type WR stars
analyzed so far have $T_{*}$ in the range $\sim 120-150kK$ (see Koesterke
2001). Regarding six LMC WC4 (Pop. I) stars and using CMFGEN, Crowther 
et al. (2002) found on average $T_{*} \sim 87kK$ and $T_{eff} \sim 70kK$, 
more in line with our values. This point will be discussed further 
in Section 5. 

The parameters obtained are not so different from those of NGC 40, which is a [WCL].
The very different spectral appearance in this case relies on the terminal 
velocity, which is $1000$ km $s^{-1}$ for NGC 40. Values up to $3600$ km $s^{-1}$
as determined by Feibelman (1998) were tested and are too high, making several 
features much broader than the observed. 

The adopted final chemical abundance is : $\beta_{He} = 43\%$, $\beta_{C} =
51\%$ and $\beta_{O} = 6\%$ (by mass), which corresponds to C/He = 0.40 and
O/He = 0.035 (by number). This result will be discussed further in Section 5. 
For the determination of O/He we used the $\lambda 5595$ feature, which is 
due to \ion{O}{3} $\lambda 5592$ and some \ion{O}{5} transitions that have $\lambda 5598$ as the 
main contributor. Again, no trace of nitrogen was found in the observed
spectra. If present, its abundance is $\beta_N < 0.1\%$ and it does not
produce any conspicuous transitions. 

Regarding iron, we could see an improvement in the fit of the region
$1400-1500$\AA\, using an abundance lower than solar. This is shown in
Fig. 9. We estimate the following range for 
this object : $ 0.1 \lesssim \beta_{Fe}/\beta_{Fe_{\odot}} < 1$. 
This is another result supporting iron depletion by neutron capture 
occurring during and after the AGB phase. As in the case of BD +30 3639, 
we could not obtain a reliable value for the Fe/Ni ratio. For NGC 5315, 
the inclusion of nickel did not produce visible transitions in the spectrum.

%
% iron : e34, e40 and e40b
%

%
% remember that the 2900-3200 lines does not allow a comparison
% because the continuum in this region is 
% underestimated... nebular contamination !
%
% e42 : C/He=0.68   O/He=0.03    C/He/O = 32/64/4
% e41 : C/He=0.25   O/He=0.03    C/He/O = 53/40/7
% these two models are based on e34
%
% e44 : O/He=0.049    C/He=0.40
% e45 : O/He=0.021    C/He=0.40
% these two models are based on e40 

\subsection{NGC 6905}

NGC 6905 is the hottest central star of our sample. Crowther et al. (1998) 
assign a [WO1] classification for this object while the [WO2] type is
preferred by Acker \& Neiner (2003). These two classes represent the highest 
ionization stages among [WR] stars. This is readily seen in its spectrum 
through intense and broad emission lines of \ion{O}{6}, \ion{C}{4} and \ion{He}{2}. 

The first spectroscopic study of this object, based on non LTE expanding 
atmosphere models, was presented by Koesterke \& Hamann (1997b). Again, line-blanketing 
and clumping were not considered. Pe$\tilde{n}$a et al. (1998) 
extended this work by analyzing the planetary nebula through a photoionization 
code. With the stellar model ionizing flux as an input parameter, the
ionization structure and electronic temperature of the nebula were reproduced 
reasonably well, but some important discrepancies remained. 
For example, the H$\beta$ theoretical flux obtained was less than 
observed, which suggests the need for better central star parameter 
estimates and up to date physics. 

The IUE spectra of NGC 6905 were studied qualitatively by Feibelman (1996). 
Several useful line identifications and flux measurements from both low and
high dispersion spectra were presented and two high terminal velocities 
were derived : 3800 km $s^{-1}$ from \ion{C}{4} $\lambda 1549$ and 2700 km
$s^{-1}$ from \ion{O}{5} $\lambda 1371$.

\subsubsection{Optical and ultraviolet fit}

In Fig. 10 we present the observed and theoretical optical spectra. 
The derived distance and reddening are d $= 1.75$kpc and E(B-V) $= 0.20$. An important
transition in this region is \ion{O}{6} $\lambda \lambda 3811, 34$. 
Previous early-type [WR] models had difficulty reproducing the strength of this line,
generally predicting a profile weaker than observed by a factor of $\sim
2$ (see Pe$\tilde{n}$a et al. 1998). The fit achieved by us, although somewhat weaker
than the observed, it is reasonable and it results from our compromise with other \ion{O}{6}
lines and \ion{O}{5} $\lambda 1371$ regarding temperature changes. The feature in 
$\sim 4660$\AA\, results from added contributions of \ion{C}{4} (mainly $\lambda 4658$ and $\lambda 4684$) 
and \ion{He}{2} ($\lambda 4686$). Other less intense lines are : \ion{O}{6}
$\lambda 5291$, \ion{He}{2} $\lambda 5412$, \ion{C}{4} $\lambda 5471$ and 
\ion{C}{4} $\lambda \lambda 5801, 12$. Contrary to the other stars in this
work, given its high temperature, NGC 6905 does not present \ion{C}{3}. 

In Fig. 11 (top) we present the far-ultraviolet fit to the FUSE spectra. 
We identified : \ion{O}{6} $\lambda \lambda 1032, 38$, $\lambda 1081$ and
$\lambda 1125$ and \ion{C}{4} $\lambda 1108$ and $\lambda 1169$. The oxygen lines
were used together with \ion{O}{5} $\lambda 1371$ to constrain the effective temperature.
In order to better fit the continuum, a slightly lower value for E(B-V) of 0.15 had to be used.

We used low resolution IUE spectra for the $\sim 1000-3000$\AA\, interval. Our
fit is also shown in Fig. 11. Between $1200-2000$\AA\, we have three stellar 
lines : \ion{O}{5} $\lambda 1371$, \ion{C}{4} $\lambda 1549$ and \ion{He}{2}
$\lambda 1640$. Helium lines are usually well reproduced in our models, 
but clearly this is not the case of \ion{He}{2} $\lambda 1640$. This line has
a contamination by the nebula and as Feibelman (1996) highlighted, its
intensity might change with the position angle of the IUE aperture. 
The remaining feature, at $1909$\AA, is a nebular \ion{C}{3}
semi-forbidden transition. 

We found some differences between the predicted and observed spectrum 
in the $2000-3000$\AA\, interval. The only theoretical lines that get 
close to the observed ones are \ion{O}{6} $\lambda 2070$, \ion{C}{4} $\lambda 2529$ and 
$\lambda 2905$. Feibelman (1996) claimed that several lines in this IUE region 
are due to \ion{Fe}{2}, \ion{O}{3}, [\ion{Ar}{5}], [\ion{Mg}{5}] and
[\ion{Ne}{4}\ion{-}{5}]. Taking into account that \ion{Fe}{2} and \ion{O}{3} 
transitions are not seen in our models (due to the high temperature) and that 
some features are related to forbidden transitions, we suggest that at least 
part of the discrepancies seen have a nebular origin. Lines from highly ionized 
elements such as \ion{C}{5}, \ion{C}{6}, \ion{O}{7} and \ion{O}{8} (not
included in our models) might also contribute to the emissions in this 
region. They are present in the spectrum of some [WR] stars and also 
in some PG 1159 stars and DO white dwarfs (see for example Koesterke \& Hamann 1997a).

We considered the possibility that an enhanced neon abundance could fit the 
emission near 2200\AA\,. An oversolar amount of this element is 
expected from evolutionary calculations and was indeed necessary to fit the spectra 
of some CSPN (Herald et al. 2005). We tested neon abundances up to 10$\times$
the solar value and the resultant \ion{Ne}{6} lines did not have the right wavelengths 
to adequately match the observed profile. We emphasize that this is not the
only unidentified emission in the spectrum of NGC 6905. In the optical for example, our
model could not reproduce a weak, broad emission at 4920\AA. A similar stellar 
feature at $\sim 2200$\AA\,, which also remains unidentified, is seen in the 
spectrum of the hot, massive star Sand 2 (Crowther et al. 2000). 

\subsubsection{Physical parameters and chemical abundances}
% final model : /john2

The physical parameters of our final model can be found in Table 2. Compared to NGC 5315, NGC 6905
is hotter (approximately a factor of 2), and has both a smaller radius and a denser wind. Although
they both are early-type [WR] stars, their spectral appearances are quite
distinct. NGC 5315 still presents \ion{C}{3} and \ion{O}{6} is weak or absent,
whilst \ion{C}{4} and \ion{O}{6} are dominant in NGC 6905. Regarding the temperature, 
Koesterke \& Hamann (1997b) found $\sim 141kK$ and our value is $\sim 150kK$. 
As no particular distance is adopted in their work, we are only able 
to compare the transformed radius $R_{T}$, which they found to be $\sim 3R_{\odot}$ 
while we found $\sim 5R_{\odot}$ without clumping.

Our final model has the following chemical abundances : $\beta_{He} =
49\%$, $\beta_{C} = 40\%$ and $\beta_{O} = 10\%$ (by mass), which corresponds 
to C/He = 0.27 and O/He = 0.05 (by number). This result will be discussed in the 
next section. Regarding nitrogen, no conspicuous transitions can be seen in
the observed spectra. We estimate an upper limit of $\beta_N < 0.1\%$. 

Because this object is very hot, we included the dominant ionization stages of iron, 
which are \ion{Fe}{7}$-${\small\rmfamily VIII}. Some subtle changes can be seen in the FUSE
region with a lower iron abundance (e.g., $1/10$ of the solar value) but they are not enough to 
determine whether there is a deficiency or not. 

% With a solar abundance, N V $\lambda 1239$ appears and is not in the IUE
% spectrum. See n15 model.

% abundance uncertainty :
%
% n30 : C/He = 0.4    O/He = 0.05
%
% n31 : C/He = 0.6    O/He = 0.05
%
% n32 : C/He = 0.15   O/He = 0.05
%
%%%%%%%%%%%%%%%%%%%%%%%%%%%%%%%%%%%%%%%%%%%%%%%%%%%%%%%%%%%%%%%%%%%%%

\section{Discussion}

\subsection{C/He mass ratios of the early-type [WR] stars} 

%%%%%% C/He determination (average the Koesterke 2001 values to see it)
It is claimed that for early-type [WR] stars the C/He mass ratio
($\beta_{C}/\beta_{He}$) is considerably smaller than in other
classes ([WCL], [WELS] and PG 1159). In Table 4 we show the average mass 
fractions of He, C and O for all H deficient central stars analyzed so far. 
The range of values measured by atmosphere models is shown for each element. 
This problem was first mentioned by Koesterke \& Hamann (1997a,b) and contradicts the evolutionary 
sequence, because early-type [WR] stars are supposed to be descendents of the [WCL] stars.
Furthermore, $\beta_{C}/\beta_{He}$ is expected to be about unity from stellar evolution 
models regardless which thermal pulse (AFTP, LTP or VLTP) is responsible for the
hydrogen deficiency (Herwig 2001). It is still unclear if such a discrepancy
is due to effects not taken into account in older models (e.g., line-blanketing). 

The determination of $\beta_{He}$ and $\beta_{C}$ is usually based on 
the transitions \ion{He}{2} $\lambda 5412$ and \ion{C}{4} $\lambda 5471$.
These lines are adopted since both are formed by recombination in a similar
region of the stellar wind. In Fig. 12 we have a spectral interval comprising these two lines for 
the two early-type stars of our sample, NGC 5315 and NGC 6905, respectively. 
In order to investigate the low $\beta_{C}/\beta_{He}$ issue found in previous works, 
we plot two models along with the observations : one with $\beta_{C}/\beta_{He} \sim
0.35$, which is the average value found for the early-type [WR] stars, and
another with this ratio approaching unity. For NGC 5315, none of the models present a reasonable
fit to these lines. We also tested lower values for $\beta_{C}$ and
$\beta_{He}$ at the expense of an increase in $\beta_{O}$. Although the fit showed
a better quality, some oxygen lines increased considerably (e.g., \ion{O}{3}\ion{-}{5}
$\lambda 5595$). In the case of NGC 6905, a $\beta_{C}/\beta_{He}$ ratio of $\sim 0.8$ 
is better than $\sim 0.35$ (bottom panel). This lower value 
is essentially the same as that adopted by Koesterke \& Hamann (1997). 
The ratio of $\sim 0.8$ is more in line with values found in other classes 
(see Table 4) and with what is expected from evolutionary calculations. 
It is very important to note however, that the fits are not perfect.
The determination of the continuum and a \ion{He}{2} $\lambda 5412$ 
nebular contamination (as is clear from the figure for NGC 6905) are factors that can 
complicate the analysis. At least from our results, it is not clear if  
early-type [WR] stars have smaller $\beta_{C}/\beta_{He}$ ratios. A larger sample
of objects with better signal-noise spectra should be carefully analyzed. In doing so, 
we should keep in mind a commitment with other parts of the spectrum when fitting \ion{He}{2} 
$\lambda 5412$ and \ion{C}{4} $\lambda 5471$.

\subsection{Silicon, phosphorus, sulfur and neon abundances}

The silicon features present in the spectra of [WR] stars  
deserve special attention. Stellar evolution models predict 
that silicon has essentially a solar mass-fraction in the H-He intershell 
matter in AGB stars. As a consequence, we expect to measure roughly 
a solar value, i.e., $\beta_{Si} \sim 0.07\%$, when analyzing H-deficient CSPN 
(see Werner \& Herwig 2006 and references therein). However, 
Leuenhagen \& Hamann (1998) derived $\beta_{Si} = 0.5 - 3\%$ in a couple of
[WCL] stars. These mass fractions correspond to 7 to 40 times the solar value,
indicating a huge discrepancy with theory. Motivated by this, we used silicon
lines in the spectra of BD +30 3639 and NGC 40 in order to check if the same 
high over solar abundances would be found. It turned out that the transitions 
\ion{Si}{4} $\lambda 1722$ in NGC 40 and  \ion{Si}{4} $\lambda 1394$, $\lambda 1403$ and
$\lambda 4089$ in BD +30 3639 were well reproduced using a solar silicon
abundance. Despite an uncertainty of approximately a factor of two in our determination, 
our results are in agreement with stellar evolution models. As reported by 
Werner \& Herwig (2006), near solar values were also found in a few PG 1159
stars. 

The atomic model for silicon used in the 
work of Leuenhagen \& Hamann (1998) was relatively simple (see their table 1). Moreover, 
neither clumping nor metal line-blanketing was taken into account 
in their study. Therefore, it would be interesting to re-analyze 
the stars in their sample in order to check if more sophisticated 
models could influence the silicon abundance determination.
This could confirm or nullify the conflict with stellar evolution models.

%
% Uncertainty : Increasing $\beta_{Si}$ up to $\sim 0.3\%$ (four times solar) 
% make these transitions more intense than the observations !
% see ~/models/ngc40/silicon2/ and ~/models/ngc40/n32 
%

Thanks to the FUSE spectra available for all stars in our sample, we could 
identify phosphorus in the spectrum of BD +30 3639, NGC 40 and NGC 5315. The
$1100-1130$\AA\, interval could only be fitted after the inclusion of \ion{P}{5}, 
which gave origin to the transitions $\lambda 1118$ and $\lambda 1128$ 
($3p\,\,^{2}P^{o}-3s\,\,^{2}S$). From nucleosynthesis calculations, phosphorus is 
expected to be observed with $4-25$ times the solar abundance, depending on 
the treatment of convective mixing (Werner \& Herwig 2006). In fact, 
the NGC 5315 spectrum could be better matched using abundances in this range. 
On the other hand, BD +30 3639 and NGC 40 models were too insensitive to 
changes in the phosphorus abundance for a reliable abundance
to be derived. They allow not only over solar values in 
the expected range but also a solar abundance. Hence, we can say that 
the result for NGC 5315 supports nucleosynthesis predictions, while for
the other two stars the error in the abundance determination is too
large to draw any conclusions.

% models/ngc5315/phosp1, models/ngc5315/phosp2 and etc. (the same for the
% other stars)

Evolutionary calculations predict that sulfur should 
be observed with an abundance of $0.6-0.9$ of the solar value 
(Werner \& Herwig 2006). We searched for sulfur lines in the 
spectra of each star of our sample. In all objects they are  
very weak or undetectable. From our models, we found that 
the following lines can be used as diagnostics for the abundance 
determination : \ion{S}{5} $\lambda 1502$, \ion{S}{5} $\lambda 2654$, and 
\ion{S}{4} $\lambda 3107$. For BD +30 3639 we estimate an upper limit 
of $\beta_S \lesssim \beta_{S_{\odot}}$, in agreement with what is 
expected from stellar evolution models. For NGC 40 and NGC 5315, a 
higher limit was adopted, namely, $\beta_S \lesssim
2\beta_{S_{\odot}}$. We could not estimate the sulfur abundance in 
NGC 6905. Given its high temperature, \ion{S}{6} is fully ionized 
throughout its stellar wind and we currently lack \ion{S}{7} atomic data.

Neon is expected to be overabundant in hydrogen deficient CSPN. According 
to Werner \& Herwig (2006), its mass frass fraction should be $\sim 2\%$,
i.e., about 10 times the solar value, although slightly lower or higher 
values are predicted depending on the initial mass of the star. 
In our models, neon diagnostic lines were found in the 2000-3000\AA\, interval. For NGC 40 and NGC 5315, 
the IUE spectra in this region have low resolution, limiting the 
accuracy of the analysis. No conspicuous neon transitions are seen. From the lines 
\ion{Ne}{4} $\lambda 2364$, \ion{Ne}{3} $\lambda 2553$ and \ion{Ne}{3}
$\lambda 2678$, we estimate a rough upper limit of $\sim 8$ times the 
solar neon value. For NGC 6905, by using the theoretical transitions \ion{Ne}{6} $\lambda 2641$ and 
\ion{Ne}{6} $\lambda 2687$ we estimate an upper limit of 
5 times the solar neon value. Higher abundances make these 
lines much stronger than observed.

In contrast to the other stars, BD +30 3639 present neon lines 
in the near UV. Although the IUE spectrum in the 2000-3000\AA\, 
region does not have a good signal-noise ratio, we can identify 
\ion{Ne}{3} $\lambda 2553$ and \ion{Ne}{3} $\lambda 2678$ 
(see Fig. 3). With a solar neon abundance, this last transition 
is practically absent in the models. We could only have a reasonable agreement 
with the observed \ion{Ne}{3} $\lambda 2678$ line by using a neon 
mass fraction of $\sim 2\%$. Therefore, BD +30 3639 is the only case 
in our sample that definitely support the prediction of stellar 
evolution models.

\subsection{Evolutionary sequence}
\label{sect_ev_seq}

A comparison among our results and previous works that did not include 
line-blanketing and clumping can be seen in Fig. 13. In this diagram, the suggested evolution is 
to decrease the transformed radius while the temperature increases when 
going from the [WCL] to the early-type [WR] stars. This would be the result of 
a shrinking radius and higher mass-loss. As we get near the white dwarf phase,
i.e.,  shortly afterwards the [WELS] and PG 1159 stage, a wind shut down is
expected (Koesterke \& Werner 1998). The abrupt decrease in
mass-loss rate increases the transformed radius. At the same time, regarding 
the spectra, the number of absorptions starts to overwhelm the number of 
emission lines. From our results, we can conclude that : 

\begin{itemize}

\item The inclusion of clumping in [WR] models can change the 
      $R_T \times T_*$ diagram in Fig. 13 considerably. If clumping is present in all [WR] stars in 
      a uniform way, i.e., if they can be described by the same filling 
      factor $f$, the mass-losses will be diminished by a factor 
      of $\sim 3$ (for $f=0.1$). Thus, we will have a vertical 
      shift of the same amount for each star because the transformed radius 
      is inversely proportional to $\dot{M}^{2/3}$ :

\[  R_T = R_* \left( \frac {v_\infty/2500 km\, s^{-1}}
    {\dot{M}/10^{-4}M_{\odot}\,yr^{-1}}\right )^{2/3}.  \]

      Among the analyzed stars, direct evidence for clumping is shown only for 
      BD +30 3639 and NGC 40. 

\item Compared to previous homogeneous models, the transformed radii increase
  to $\sim 0.5$ dex. Without clumping, we still found a change of $\sim
  0.1-0.2$ dex. 
  
\item Our result for NGC 6905 suggests that the gap found between early-type 
      [WR] stars and [WELS] could disappear if clumping is not important or 
      present in this latter group. 

\item A far-UV to optical coverage plus the use of line-blanketing is probably 
    the reason for the change in the temperatures of all stars regarding 
    previous results. However, we did not see any systematic change that 
    would allow any conclusion regarding the evolutionary sequence.
    We decreased the temperatures of NGC 40 and 
    NGC 5315 and increased the temperatures of NGC 6905 and BD +30 3639. An important
    case to note is the one of NGC 5315, which will be discussed in the next
    section. This star is clearly far from the other early-type objects. 
    An analysis of a larger sample is required to check if the same behavior can be found. 

\end{itemize}

Fig. 13 allows us to check the differences between our work and previous
results. However, it is interesting to compare our results also with  
evolutionary tracks in the HR diagram. Indeed, this kind of comparison 
was one of the strongest arguments first supporting an evolutionary connection 
between [WR], [WELS] and PG 1159 stars (see Hamann 1997). Unfortunately, 
there are some drawbacks in doing this. First, the radii determined 
from the models are distant dependent whilst Fig. 13 is not. Hence, the 
determination of surface gravities can be very uncertain. Absorption 
lines in PG 1159 stars are used to overcome this difficulty, but the same 
is not possible for the [WR] stars. Another concern are the evolutionary
tracks. Generally, the ones used are tailored for H-rich objects, which could 
be another source of uncertainty. The main reason for this is the complexities 
involved in getting a H-deficient object and its evolution in an appropriate
way. Nevertheless, very recently, Althaus et al. (2005) made available a complete evolutionary 
track for a 2.7 $M_{\odot}$ star from the main sequence up to the normal (H-rich) white 
dwarf stage. Further, after a very late thermal pulse (VLTP), the calculations
were carried out to the H-deficient PG 1159 stage and finally, to the  DO, DB
and DQ white dwarf region. Their work fits our needs and is reproduced here in our
Fig. 14. The number of data points are augmented by our analysis and by new 
results regarding PG 1159 stars (H\"ugelmeyer et al. 2005). We also add to 
the plot all the previously analyzed Galactic [WR] stars (Koesterke 2001). 
All [WR] stars are assumed to have 0.6$M_{\odot}$. 
Although the [WR] $\rightarrow$ PG 1159 evolution can be inferred from Fig. 14 
(based on an appropriate evolutionary track), some remarks should be made :

\begin{itemize}

\item There is not a smooth transition from [WCL] to the early-type [WR] stars
  (see Log $T_{*} > 4.9$). This reflects the long standing issue regarding  
  the lack of stars in the subtypes [WC5-7], i.e., there are no [WR] 
  stars with $T_{*}$ between $\sim 80kK$ and $\sim 110kK$;

\item NGC 5315 ([WC4]-[WO4]) is clearly displaced from other early-type [WR] 
  stars. The majority of early-type [WR] stars are distributed over a very narrow region near 
  the PG 1159 stars, where $T_{*}$ $> 100kK$. In contrast, the position of 
  NGC 5315 is practically the same as NGC 40, which is a [WCL] star. 
  It must be kept in mind that although some early-type [WR] and PG 1159 stars  
  occupy a region in common in this diagram (near Log g = 6), their spectra 
  are quite distinct. In contrast with PG 1159 stars, all early-type [WR] 
  stars present mass-loss;

\item As already noted by Hamann (1997), the position of [WELS] in
    Fig. 14 is slightly before the early-type [WR] stars. This is another point 
    showing the need for the analysis of more objects in this class.

\end{itemize}

Until we elucidate the points described above, the detailed evolutionary 
sequence [WCL] $\rightarrow$ [WCE] $\rightarrow$ [WO] $\rightarrow$ [WELS]
$\rightarrow$ PG 1159 must be considered with caution.

It would be very useful to have the same diagram but for the other two thermal
pulse models (AFTP and LTP) and for different remnant stellar masses. This is 
beyond the scope of the present work.

\subsection{The early-type [WR] star NGC 5315}

As already mentioned, the $T_{*}$ found for NGC 5315 is lower than 
previous determinations for other [WC4] stars, namely : IC 1747, NGC 1501 and 
NGC 6369 (see Koesterke 2001). However, it is important to note that they all 
now have different stellar classifications. Recently, because some [WCE] stars 
present oxygen lines more intense than carbon lines, new classification 
systems (Crowther et al. 1998; Acker \& Neiner 2003) replaced [WC2-3] by [WO1-4] 
subtypes, changing the status of several objects (including some [WC4]) and 
giving a more appropriate description of the ionization of the wind. It turns out that 
NGC 5315 now is the only object analyzed from its class, which according to Crowther et
al. (1998) is [WC4] and to Acker \& Neiner (2003), [WO4]. As a matter of fact, 
NGC 5315 is clearly separated from other early-type [WR] stars in the diagrams in
Figs. 13 and 14. This result should not be interpreted as the influence of the use of 
metal line-blanketing, and calls for an analysis of more objects in the
[WC4]-[WO4] class.

\section{Summary}

We presented detailed far-UV to optical analysis of four [WR] stars.
New physical parameters and abundances for the central stars BD +30 3639, 
NGC 40, NGC 5315 and NGC 6905 were obtained using CMFGEN. Contrary to 
most of the previous works about these objects, metal line-blanketing and 
a simple treatment of clumping were taken into account. The early-type [WR] 
star NGC 5315 was studied for the first time by means of non LTE expanding 
atmosphere models. So far, no other object from its spectral class
([WC4]-[WO4]) has been analyzed with a similar code.

Far-UV FUSE observations of all stars were analyzed for the first time. 
Phosphorus was found in the spectra of BD +30 3639, NGC 40 and NGC 5315. 
The transitions $\lambda 1118$ and 
$\lambda 1128$ could be identified as \ion{P}{5} ($3p\,\,^{2}P^{o}-3s\,\,^{2}S$). 
Regarding its abundance, the result for NGC 5315 agrees with 
nucleosynthesis calculations during the AGB phase, where a range of 
$4-25$ times the solar value is predicted. For the other objects, the models 
were too insensitive to abundance changes and a solar value is also possible. 

We derived a solar silicon abundance for two stars of our sample : 
BD +30 3639 and NGC 40. Although we estimate an uncertainty of a factor of 
two in this determination, our result is in agreement with the prediction 
that the silicon abundance should not show a significant deviation 
from the solar value.

We estimated upper limits for the amount of sulfur in BD +30 3639, NGC 40 
and NGC 5315. Regarding neon, upper limits for its abundance were established 
for NGC 40, NGC 5315 and NGC 6905. In order to provide a reasonable fit 
to the \ion{Ne}{3} $\lambda 2678$ line in BD +30 3639, we used  
a neon abundance of $\sim 2\%$ (mass fraction). This result supports 
the prediction of an oversolar neon abundance in hydrogen 
deficient CSPN.

An analysis of the iron abundance was performed for each star. 
Evidence of depletion was found for BD +30 3639 and NGC 5315, while 
for the other stars, the spectrum was not sensitive enough to iron abundance
changes to allow a reliable estimate. The confirmation of a depletion supports theoretical 
calculations based on neutron capture during and after the AGB phase. 
A parallel confirmation of iron depletion through the enrichment of 
nickel could not be achieved for BD +30 3639 and NGC 5315. At present, it is 
very difficult to determine a reliable abundance for this element.

Using the results for NGC 5315 and NGC 6905, we addressed the issue of the 
low C/He mass ratios of the early-type [WR] stars. We had difficulties in 
obtaining a simultaneous fit of \ion{He}{2} $\lambda 5412$ and \ion{C}{4} $\lambda 5471$ 
in the models for NGC 5315. Neither the typical low value of $\beta_C/\beta_{He} \sim
0.35$ found in previous works or $\beta_C/\beta_{He} \sim 1$ was able to 
satisfactorily reproduce these lines. For NGC 6905, a value of $\beta_C/\beta_{He} \sim 0.8$ 
is better than a lower one and it is more compatible with values found 
in the other classes. For a more reliable analysis, we need spectra with good flux
calibrations and high S/N. In this way, we can better determine the stellar 
continuum, better measure the strengths and profiles of weak
features, and better separate the nebular contamination. 

The impact of our results on the evolutionary sequence [WR] $\rightarrow$ PG
1159 is shown in the transformed radius-temperature and HR diagrams. The
latter made use of the evolutionary track of Althaus et al. (2005), which is 
suitable for H-deficient objects. Although there are displacements regarding 
previous results in these diagrams due to the new physics added (e.g., metal
line-blanketing and clumping), a deeper comprehension of the evolution of
these stars requires the investigation of a larger sample. In particular, 
an analysis of more stars of the [WELS] class and of other early-type [WR] 
stars with the same spectral classification as NGC 5315 is desirable.

%%%%%%%%%%%%%%%%%%%%%%%%%%%%%%%%%%%%%%%%%%%%%%%%%%%%%%%%%%%%%%%

\begin{acknowledgements}

W. M. thanks Funda\c c\~ao de Amparo \`a Pesquisa do Estado do Rio de Janeiro (FAPERJ) 
for financial support. DJH gratefully acknowledges partial support from NASA 
LTSA grant NAG5-8211.

\end{acknowledgements}

{}

\clearpage
\thispagestyle{empty}
\setlength{\voffset}{-15mm}
\begin{table}[!p]
\caption{Atomic Model for BD +30 3639.}
\begin{tabular}{|l|c|c|}\hline
%%%%%%%%%%%%%%%%%%%%%%%%%%%%%%%%%%%%%%%%%%%%%%%%%%%%%%%%%%%%%%%%%%%%%%%%%%%%%%%%%%

Ion        & Number of full levels &  Number of super-levels \\\hline

%%%%%%%%%%%%%%%%%%%%%%%%%%%%%%%%%%%%%%%%%%%%%%%%%%%%%%%%%%%%%%%%%%%%%%%%%%%%%%%%%%%
\ion{He}{1}       &  39  & 27  \\\hline
\ion{He}{2}      &  30  & 13  \\\hline
\ion{C}{2}       & 338  & 104 \\\hline
\ion{C}{3}      & 243  & 99  \\\hline
\ion{C}{4}       &  64  & 49  \\\hline
\ion{O}{2}       & 111  & 30  \\\hline
\ion{O}{3}      & 349  & 267 \\\hline
\ion{O}{4}       &  72  & 30  \\\hline
\ion{O}{5}       &  91  & 31  \\\hline
\ion{O}{6}       &  19  & 13  \\\hline
\ion{Ne}{2}      &  48  & 14  \\\hline
\ion{Ne}{3}     &  71  & 23  \\\hline
\ion{Ne}{4}      &  52  & 17  \\\hline
\ion{Mg}{2}      &  45  & 18  \\\hline
\ion{Al}{2}      &  58  & 38  \\\hline
\ion{Al}{3}     &  45  & 17  \\\hline
\ion{Si}{2}      &  80  & 52  \\\hline
\ion{Si}{3}     &  45  & 25  \\\hline
\ion{Si}{4}      &  38  & 27  \\\hline
\ion{P}{5}        &  62  & 16  \\\hline
\ion{S}{3}      &  14  & 11  \\\hline
\ion{S}{4}       &  23  & 19  \\\hline
\ion{S}{5}        &  22  & 21  \\\hline
\ion{S}{6}       &  19  & 17  \\\hline
\ion{Ar}{3}     &  36  & 10  \\\hline
\ion{Ar}{4}      &  61  & 19  \\\hline
\ion{Ar}{5}       &  36  & 18  \\\hline
\ion{Ca}{2}      &  46  & 17  \\\hline
\ion{Ca}{3}     & 110  & 33  \\\hline
\ion{Ca}{4}      & 193  & 34  \\\hline
\ion{Ca}{5}       & 121  & 45  \\\hline
\ion{Ca}{6}      & 108  & 47  \\\hline
\ion{Fe}{3}     & 477  & 61  \\\hline
\ion{Fe}{4}      & 1000 & 100 \\\hline
\ion{Fe}{5}       & 182  & 19  \\\hline
\ion{Fe}{6}      &  80  & 10  \\\hline

\end{tabular}

%\begin{minipage}{15.5cm}
\tablecomments{NGC 40 and NGC 5315 have practically the same atomic model, the only
difference being the number of super-levels in a few ions. For NGC 6905,
\ion{C}{2}, \ion{O}{2}\ion{-}{3}, \ion{Ne}{2}, \ion{Si}{2}, \ion{S}{3} and \ion{Ca}{2} were excluded, while 
\ion{Ne}{5}\ion{-}{6}, \ion{Mg}{3}, \ion{Si}{5} and \ion{Fe}{7}\ion{-}{8} were included.}

%\end{minipage}
\end{table}
\clearpage
\setlength{\voffset}{0mm}
%%%%%%%%%%%%%%%%%%%%%%%%%%%%%%%%%%%%%%%%%%%%%%%%%%%%%%%%%%%%%%%%%%%%%%%%%%%

{\rotate
\begin{table}[p]
{\scriptsize
\caption{Basic parameters derived and comparison with previous results.}
\begin{tabular}{llllllllllll}  
\hline\hline
Star & $T_{*}$ (K) & $R_{*}/R_{\sun}$ & $T_{eff}$ (K) & 
log $\dot{M}$ & log $\dot{M}/\sqrt{f} $ & $v_{\infty}$ (km $s^{-1}$) &
$R_{T}/R_{\sun}$ & $\beta_{He}$ & $\beta_{C}$ & $\beta_{O}$ & d (kpc) \\
\hline
{\bf BD +30 3639} (this work) & \multicolumn{1}{c}{48060} & \multicolumn{1}{c}{1.0} & 
            \multicolumn{1}{c}{46720} & \multicolumn{1}{c}{-6.30} & 
            \multicolumn{1}{c}{-5.80} & \multicolumn{1}{c}{700}   & 
            \multicolumn{1}{c}{6.8 (14.6)} & 
            \multicolumn{1}{c}{43} &
            \multicolumn{1}{c}{51} & \multicolumn{1}{c}{6} &  \multicolumn{1}{c}{1.2} 
\\
Leuenhagen et al. (1996) & \multicolumn{1}{c}{47000} & \multicolumn{1}{c}{1.49} & 
            \multicolumn{1}{c}{42000} & \multicolumn{1}{c}{-} & 
            \multicolumn{1}{c}{-5.40} & \multicolumn{1}{c}{700}   & 
            \multicolumn{1}{c}{5.5 (-)}  & 
            \multicolumn{1}{c}{45} &
            \multicolumn{1}{c}{50} & \multicolumn{1}{c}{5} & \multicolumn{1}{c}{}

\\
Crowther et al. (2006) & \multicolumn{1}{c}{55000} & \multicolumn{1}{c}{0.85} & 
            \multicolumn{1}{c}{48000} & \multicolumn{1}{c}{-6.05} & 
            \multicolumn{1}{c}{-5.55} & \multicolumn{1}{c}{700}   & 
            \multicolumn{1}{c}{3.9 (8.5)} & 
            \multicolumn{1}{c}{51} &
            \multicolumn{1}{c}{38} & \multicolumn{1}{c}{10} & \multicolumn{1}{c}{}
\\
\\
{\bf NGC 40} (this work) & \multicolumn{1}{c}{73310} & \multicolumn{1}{c}{0.43} & 
            \multicolumn{1}{c}{70840} & \multicolumn{1}{c}{-6.25} & 
            \multicolumn{1}{c}{-5.75} & \multicolumn{1}{c}{1000}  & 
            \multicolumn{1}{c}{3.4 (7.4) } & 
            \multicolumn{1}{c}{43} &
            \multicolumn{1}{c}{51} & \multicolumn{1}{c}{6} & \multicolumn{1}{c}{1.4}
\\
Leuenhagen et al. (1996)  & \multicolumn{1}{c}{78000} & \multicolumn{1}{c}{0.46} & 
            \multicolumn{1}{c}{46000} & \multicolumn{1}{c}{-} & 
            \multicolumn{1}{c}{-5.40} & \multicolumn{1}{c}{1000}  & 
            \multicolumn{1}{c}{2.2 (-)} & 
            \multicolumn{1}{c}{40} &
            \multicolumn{1}{c}{50} & \multicolumn{1}{c}{10} &  \multicolumn{1}{c}{}

\\
\\
{\bf NGC 5315} (this work)  & \multicolumn{1}{c}{76420} & \multicolumn{1}{c}{0.40} & 
            \multicolumn{1}{c}{74590} & \multicolumn{1}{c}{-6.33} & 
            \multicolumn{1}{c}{-5.83} & \multicolumn{1}{c}{2400}   & 
            \multicolumn{1}{c}{6.5 (13.9)}  &  
            \multicolumn{1}{c}{43} &
            \multicolumn{1}{c}{51} & \multicolumn{1}{c}{6} & \multicolumn{1}{c}{2.5}

\\
de Freitas Pacheco et al. (1986;1993) & \multicolumn{1}{c}{82700} & \multicolumn{1}{c}{0.31} & 
            \multicolumn{1}{c}{-} & \multicolumn{1}{c}{-} & 
            \multicolumn{1}{c}{-5.83} & \multicolumn{1}{c}{2600}   & 
            \multicolumn{1}{c}{5.2 (-)}  &  
            \multicolumn{1}{c}{-} &
            \multicolumn{1}{c}{-} & \multicolumn{1}{c}{-} & \multicolumn{1}{c}{}
\\
\\
{\bf NGC 6905} (this work) & \multicolumn{1}{c}{149600} & \multicolumn{1}{c}{0.10} & 
            \multicolumn{1}{c}{146200} & \multicolumn{1}{c}{-7.15}   & 
            \multicolumn{1}{c}{-6.65}  & \multicolumn{1}{c}{1890} & 
            \multicolumn{1}{c}{4.9 (10.5)}  & 
            \multicolumn{1}{c}{49} &
            \multicolumn{1}{c}{40} & \multicolumn{1}{c}{10} & \multicolumn{1}{c}{1.75}

\\
Koesterke \& Hamann (1997)  & \multicolumn{1}{c}{141000} & \multicolumn{1}{c}{-} & 
            \multicolumn{1}{c}{-} & \multicolumn{1}{c}{-}   & 
            \multicolumn{1}{c}{-}   & \multicolumn{1}{c}{1800} & 
            \multicolumn{1}{c}{3.4 (-)} &  
            \multicolumn{1}{c}{60} &
            \multicolumn{1}{c}{25} & \multicolumn{1}{c}{15} & \multicolumn{1}{c}{}

\\ \hline

\end{tabular}

%\begin{minipage}{15.5cm}
\tablecomments{Mass loss unit is $M_{\sun}$ year $^{-1}$. Values between parentheses use
      clumping. Chemical abundances are given in mass fractions. The distances
      were derived from the models.}
%\end{minipage} 
}
\end{table}
}

%%%%%%%%%%%%%%%%%%%%%%%%%%%%%%%%%%%%%%%%%%%%%%%%%%%%%%%%%%%%%%%%%%%%%%%%%%%%%

\begin{table}[!h]
\caption{Additional physical parameters of our models.}
\begin{tabular}{lcccc}
\hline\hline

Star & Log $Q_0$          & Log $Q_1$       & Log $Q_2$        & $\eta$  \\
     &  $912\AA$ (\ion{H}{1})    & $504\AA$ (\ion{He}{1}) & $228\AA$ (\ion{He}{2}) &        \\
\hline
BD +30 3639   & \multicolumn{1}{c}{47.45} & \multicolumn{1}{c}{45.89} &
\multicolumn{1}{c}{34.28} & 11.0 (3.5) \\

NGC 40      & \multicolumn{1}{c}{47.56} & \multicolumn{1}{c}{46.91} &
\multicolumn{1}{c}{36.79} & 17.6 (5.5) \\

NGC 5315    & \multicolumn{1}{c}{47.58} & \multicolumn{1}{c}{47.15} &
\multicolumn{1}{c}{37.46} & 35.6 (11.0) \\

NGC 6905    & \multicolumn{1}{c}{47.43} & \multicolumn{1}{c}{47.30} &
\multicolumn{1}{c}{46.65} & 4.1 (1.3) \\
\hline
\label{logqs}
\end{tabular}
%\begin{minipage}{15.5cm}
\tablecomments{The ionizing flux unit is {\it photons s$^{-1}$} and the values between
      parentheses for the wind efficiency $\eta$ use clumping.}
%\end{minipage} 
\end{table}

%%%%%%%%%%%%%%%%%%%%%%%%%%%%%%%%%%%%%%%%%%%%%%%%%%%%%%%%%%%%%%%%%%%%%%%%%%%%%

\begin{table}
\caption{Average mass fractions for H deficient CSPN.}
\begin{tabular}{lcccc}
\hline\hline
Class & $\bar{\beta}_{He}$ (\%) &  $\bar{\beta}_{C}$ (\%) &  $\bar{\beta}_{O}$
(\%) & $ \bar{\beta}_{C}/\bar{\beta}_{He} $
 
\\
\hline
$[$WCL$]$ (13)   & \multicolumn{1}{c}{43 $\pm$ 3} & \multicolumn{1}{c}{50 $\pm$ 4} &
\multicolumn{1}{c}{6 $\pm$ 3} & 1.16 $\pm$ 0.12
\\
$[$WCE$]$ (11)   & \multicolumn{1}{c}{66 $\pm$ 9} & \multicolumn{1}{c}{23 $\pm$ 6} &
\multicolumn{1}{c}{11 $\pm$ 4} & 0.35 $\pm$ 0.10
\\
$[$WELS$]$ (3)   & \multicolumn{1}{c}{40 $\pm$ 7} & \multicolumn{1}{c}{44 $\pm$ 5} &
\multicolumn{1}{c}{14 $\pm$ 2} & 1.10 $\pm$ 0.23
\\
PG $1159$ (13)  & \multicolumn{1}{c}{50 $\pm$ 17} & \multicolumn{1}{c}{39 $\pm$ 14} &
\multicolumn{1}{c}{10 $\pm$ 6} & 0.78 $\pm$ 0.39
\\ \hline

\end{tabular}

%\begin{minipage}{15.5cm}
\tablecomments{Data from Koesterke (2001); Reiff et al. (2005); Dreizler \& Heber
      (1998); Kruk \& Werner (1998). The number of objects considered in each
    class is between parentheses. Hybrid PG 1159 stars were omitted.}
%\end{minipage}
\end{table}

%%%%%%%%%%%%%%%%%%%%%%%%%%%%%%%%%%%%%%%%%%%%%%%%%%%%%%%%%%%%%%%%%%%%%%%%%%%%%%%%%%%%%%%

\clearpage

\begin{figure}[!ht] 
\centering  
\includegraphics[width=16cm]{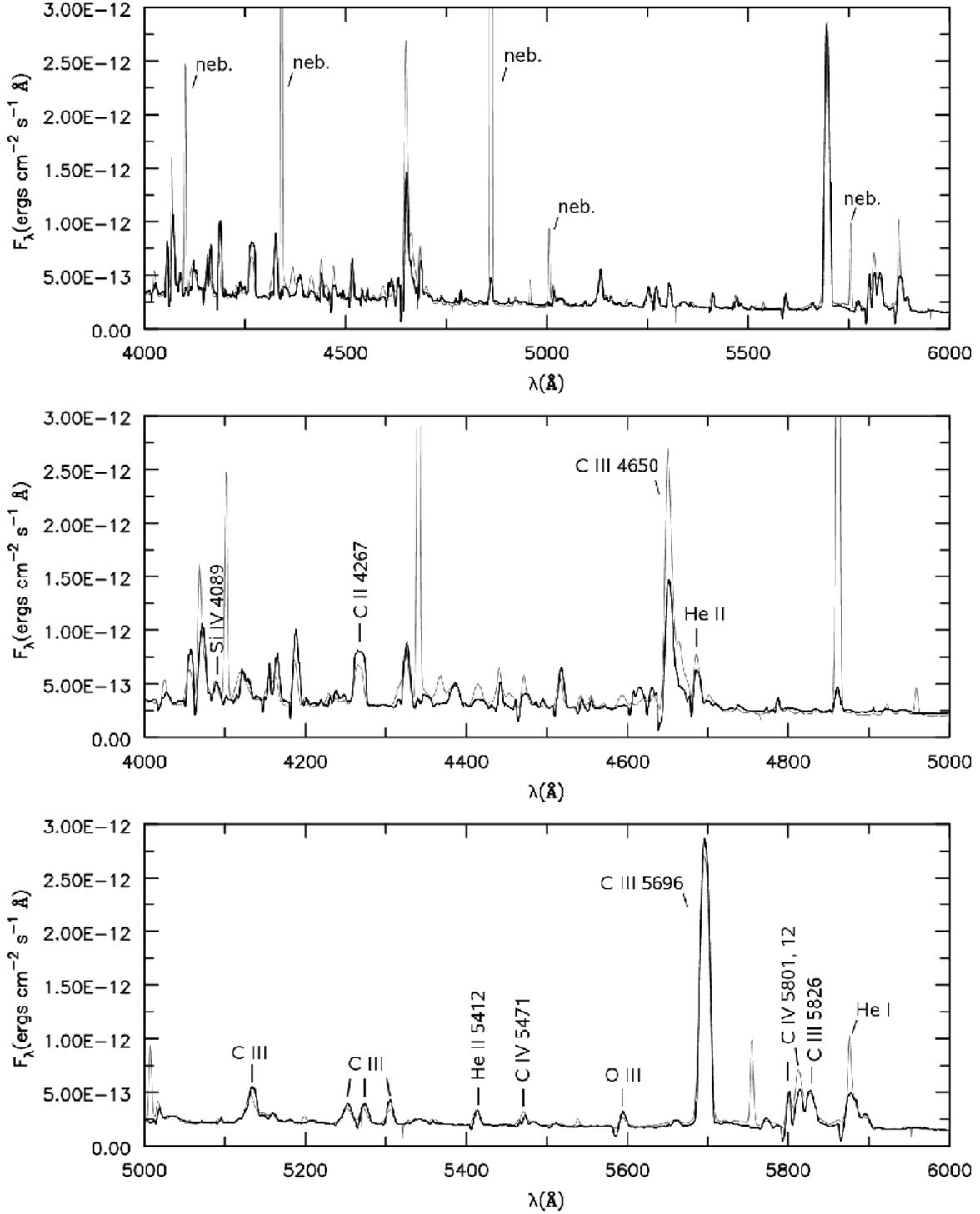}
\caption{BD +30 3639 optical spectrum and our model (thick line).
Some nebular lines are indicated in the top panel
and identified as : H$\delta$, H$\gamma$, H$\beta$,  
[\ion{O}{3}] $\lambda 5007$ and [\ion{N}{2}] $\lambda 5754$.}
\label{bd30_optical}
\end{figure} 

\begin{figure}[!ht] 
\centering  
\includegraphics[width=16cm]{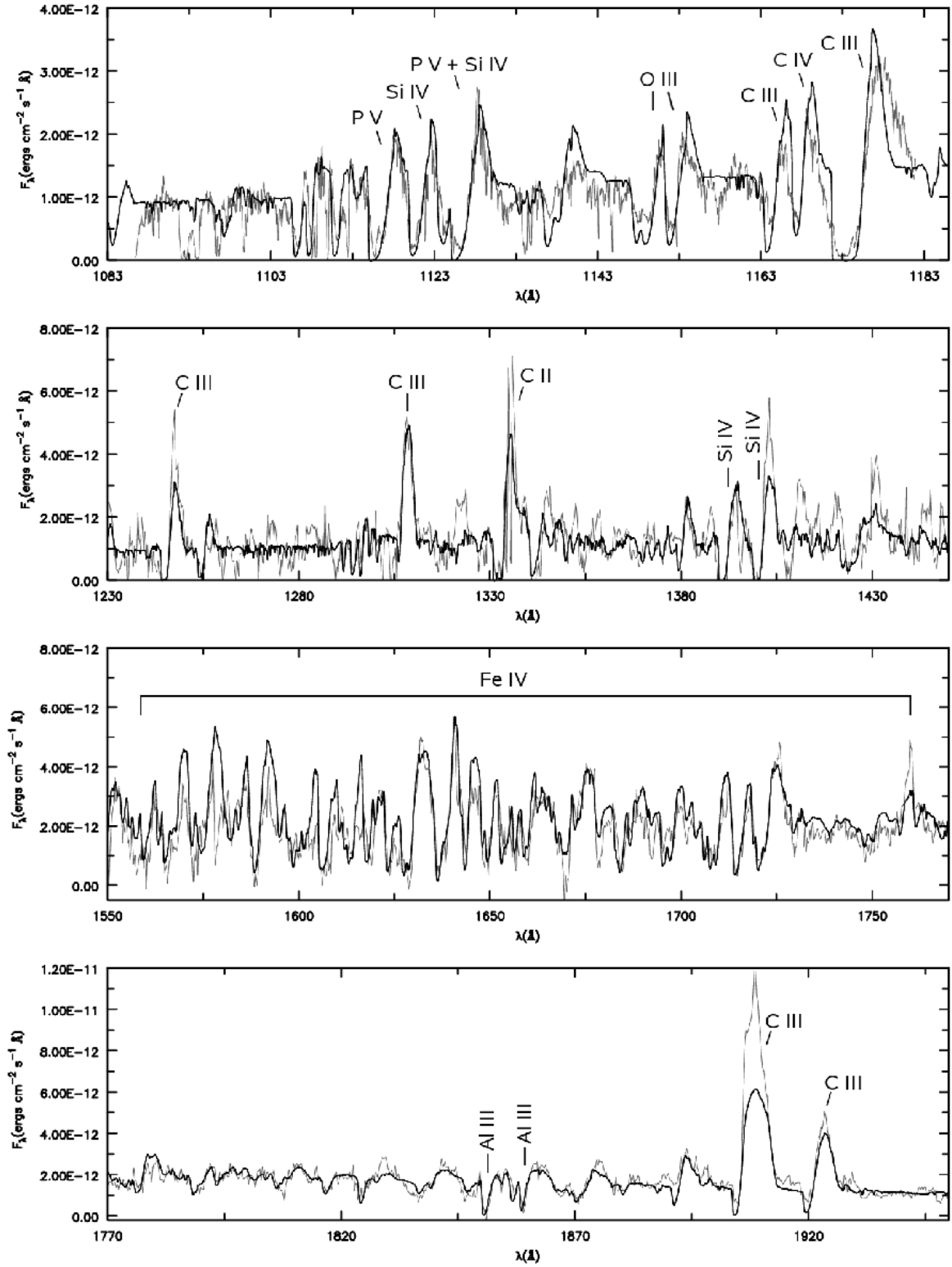}
\caption{BD +30 3639 ultraviolet spectra and our model (thick line). 
Top : FUSE spectrum. Other Panels : High resolution IUE spectrum.}
\label{bd30_uv}
\end{figure} 

\begin{figure}[!ht] 
\centering  
\includegraphics[width=16cm]{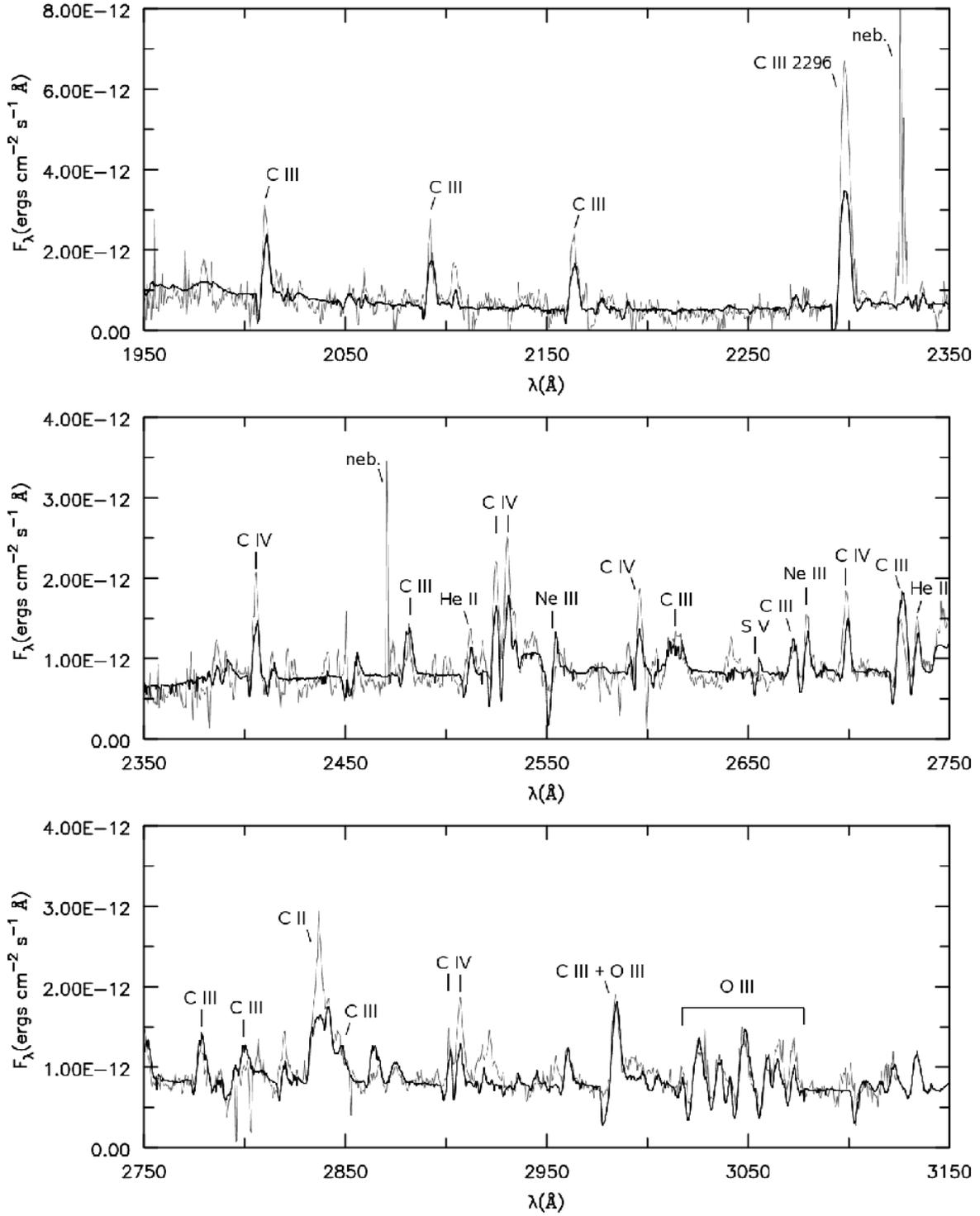}
\caption{BD +30 3639 high resolution UV spectrum and our model (thick line). 
The nebular lines beyond $2296$\AA\, are : \ion{C}{2}] ($\lambda 2323$, 
$\lambda 2325$, $\lambda 2327$, $\lambda 2328$) and [\ion{O}{2}] $\lambda 2471$.}
\label{bd30_uv2}
\end{figure} 

\begin{figure}[!ht] 
\centering  
\includegraphics[width=10cm]{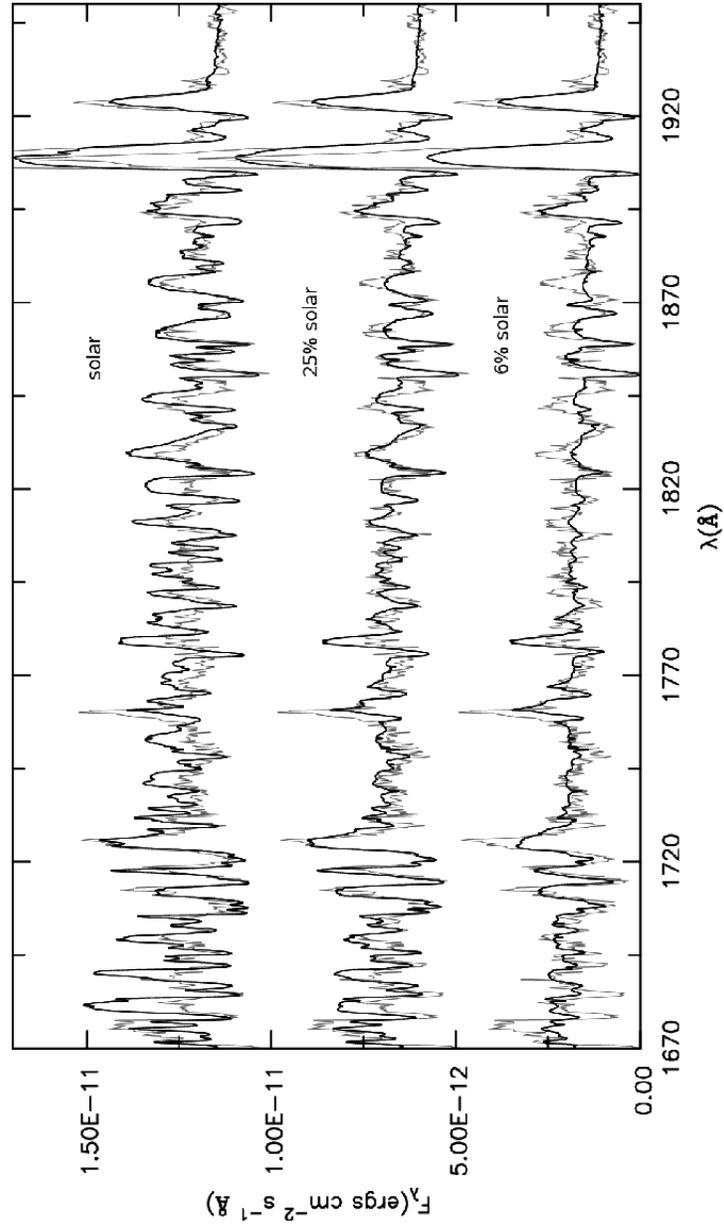}
\caption{Iron deficiency in BD +30 3639. Models are represented by thick lines.
Top : solar abundance. Middle : best fit model with 25\% of the solar
abundance. Bottom : 6\% of the solar abundance. The two upper models were 
vertically displaced for clarity. Abundances are in mass fractions.}
\label{bd30_iron}
\end{figure} 

\begin{figure}[!ht] 
\centering  
\includegraphics[width=16cm]{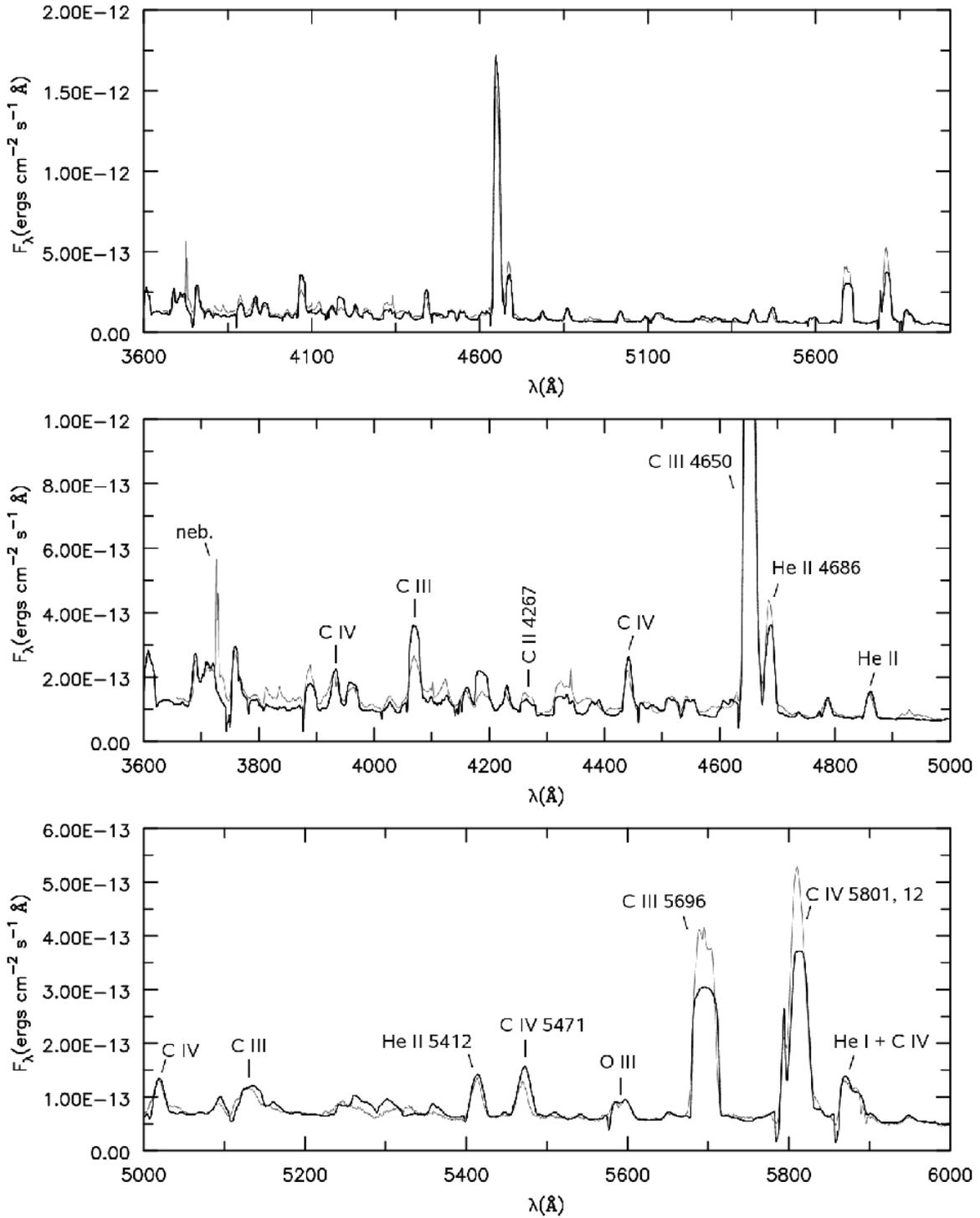}
\caption{NGC 40 optical spectrum and our model (thick line).}
\label{ngc40_optical}
\end{figure} 

\begin{figure}[!ht] 
\centering  
\includegraphics[width=16cm]{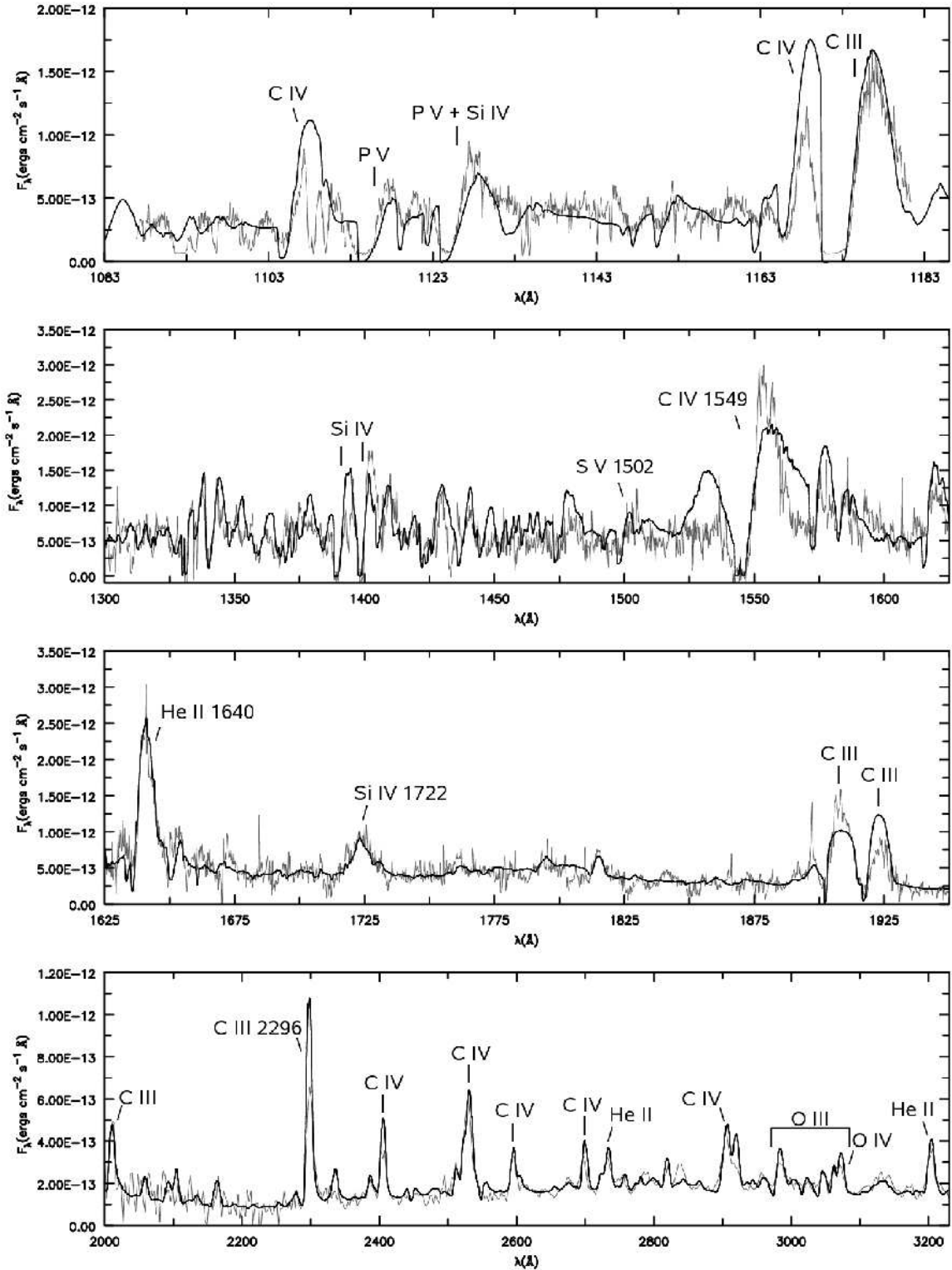}
\caption{NGC 40 ultraviolet spectra and our model (thick line). 
Top : FUSE spectrum. 2nd and 3rd panel : high resolution IUE spectrum. 
Bottom : low resolution IUE spectrum.}
\label{ngc40_uv}
\end{figure} 

\begin{figure}[!ht] 
\centering  
\includegraphics[width=16cm]{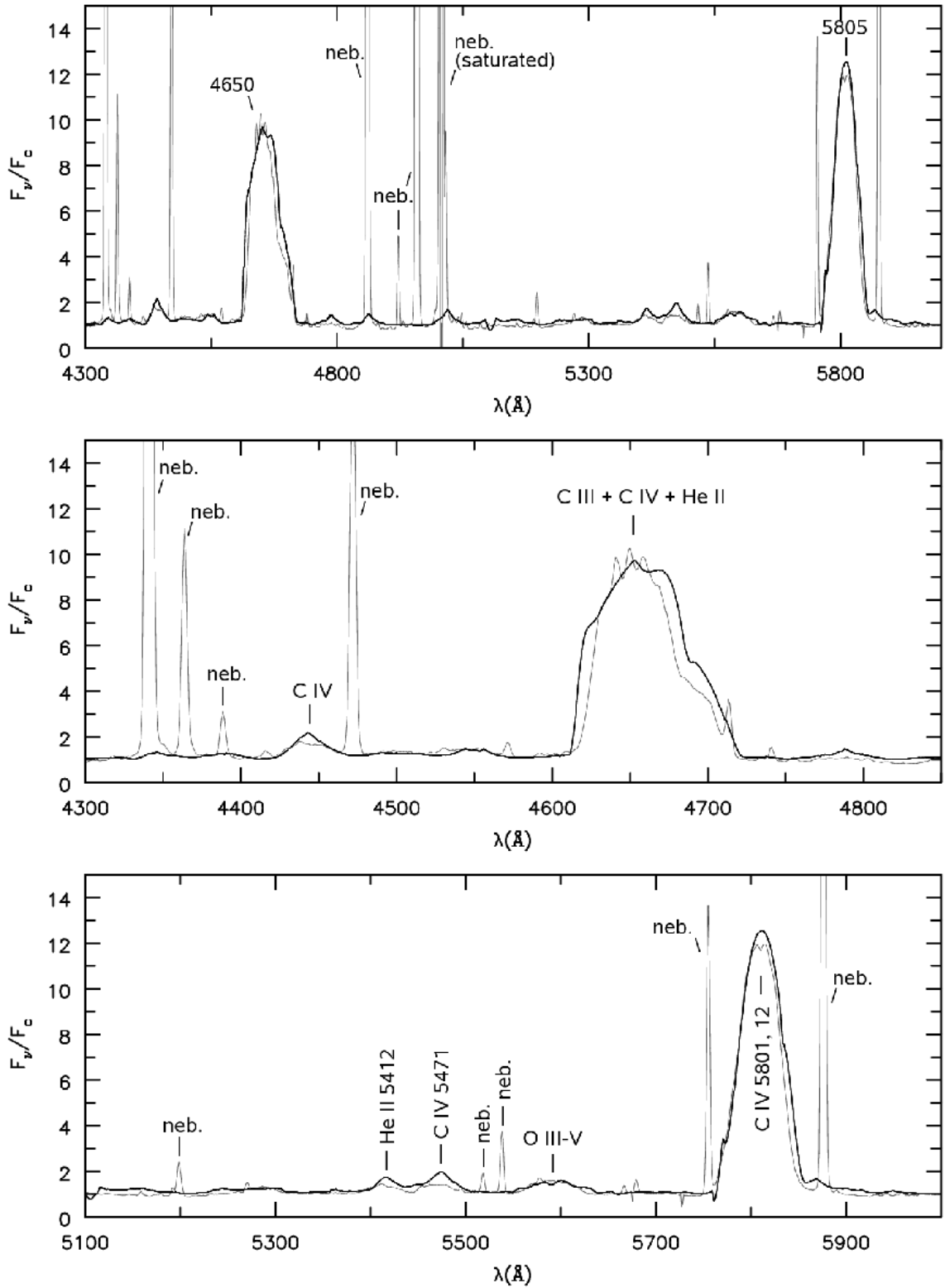}
\caption{NGC 5315 optical spectrum and our model (thick line).
The nebular lines are : H$\gamma$, [\ion{O}{3}] $\lambda 4363$, \ion{He}{1} $\lambda
4388$, $\lambda 4472$, H$\beta$, [\ion{O}{3}] $\lambda 4959$, $\lambda 5007$, 
[\ion{N}{1}] $\lambda 5199$, [\ion{Cl}{3}] $\lambda 5518$, $\lambda 5538$,
[\ion{N}{2}] $\lambda 5754$ and \ion{He}{1} $\lambda 5876$.}
\label{ngc5315_optical}
\end{figure} 

\begin{figure}[!ht] 
\centering  
\includegraphics[width=16cm]{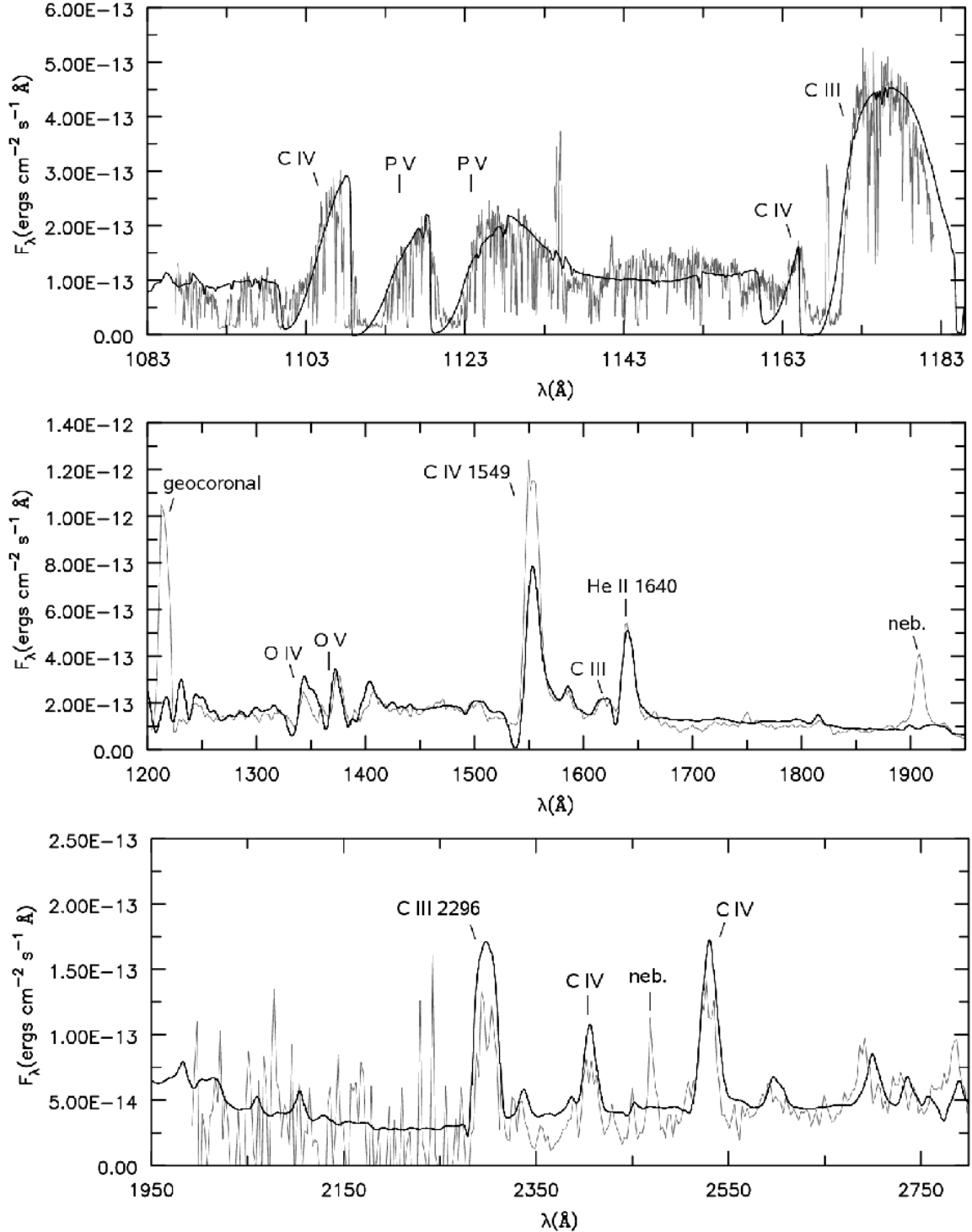}
\caption{NGC 5315 ultraviolet spectra and our model (thick line). 
Top : FUSE spectrum. Other panels : Low resolution IUE spectra.
The non-stellar lines indicated are : L$\alpha$ (geocoronal emission),  
\ion{C}{3}] $\lambda 1909$ and [\ion{O}{2}] $\lambda 2470$ (nebular).
Unusually, the observed \ion{C}{4}\ profile lacks the strong P-Cygni absorption
generally associated with this line.}
\label{ngc5315_uv}
\end{figure} 

\begin{figure}[!ht] 
\centering  
\includegraphics[width=10cm]{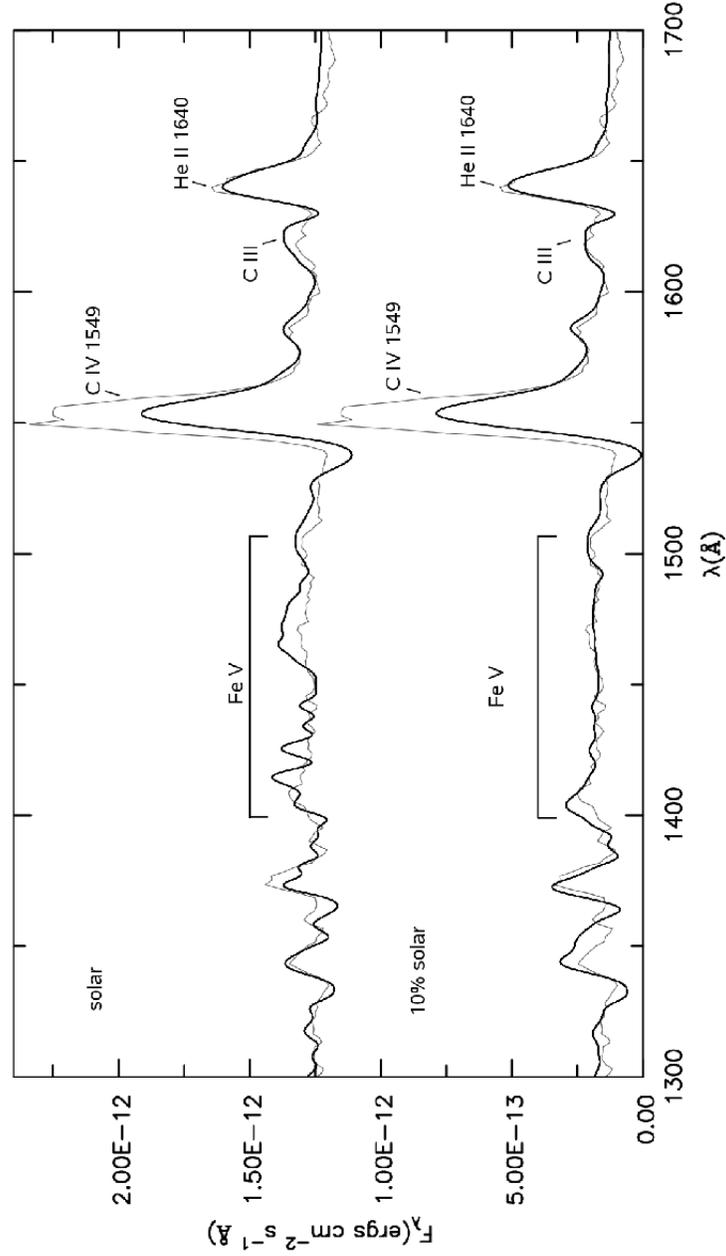}
\caption{Iron deficiency in NGC 5315. Models are represented by thick lines.
Top : solar abundance. Bottom : 10\% of the solar abundance. The upper model was
vertically displaced for clarity. Abundances are in mass fractions.}
\label{ngc5315_iron}
\end{figure} 

\begin{figure}[!ht] 
\centering  
\includegraphics[width=17cm, height=16cm]{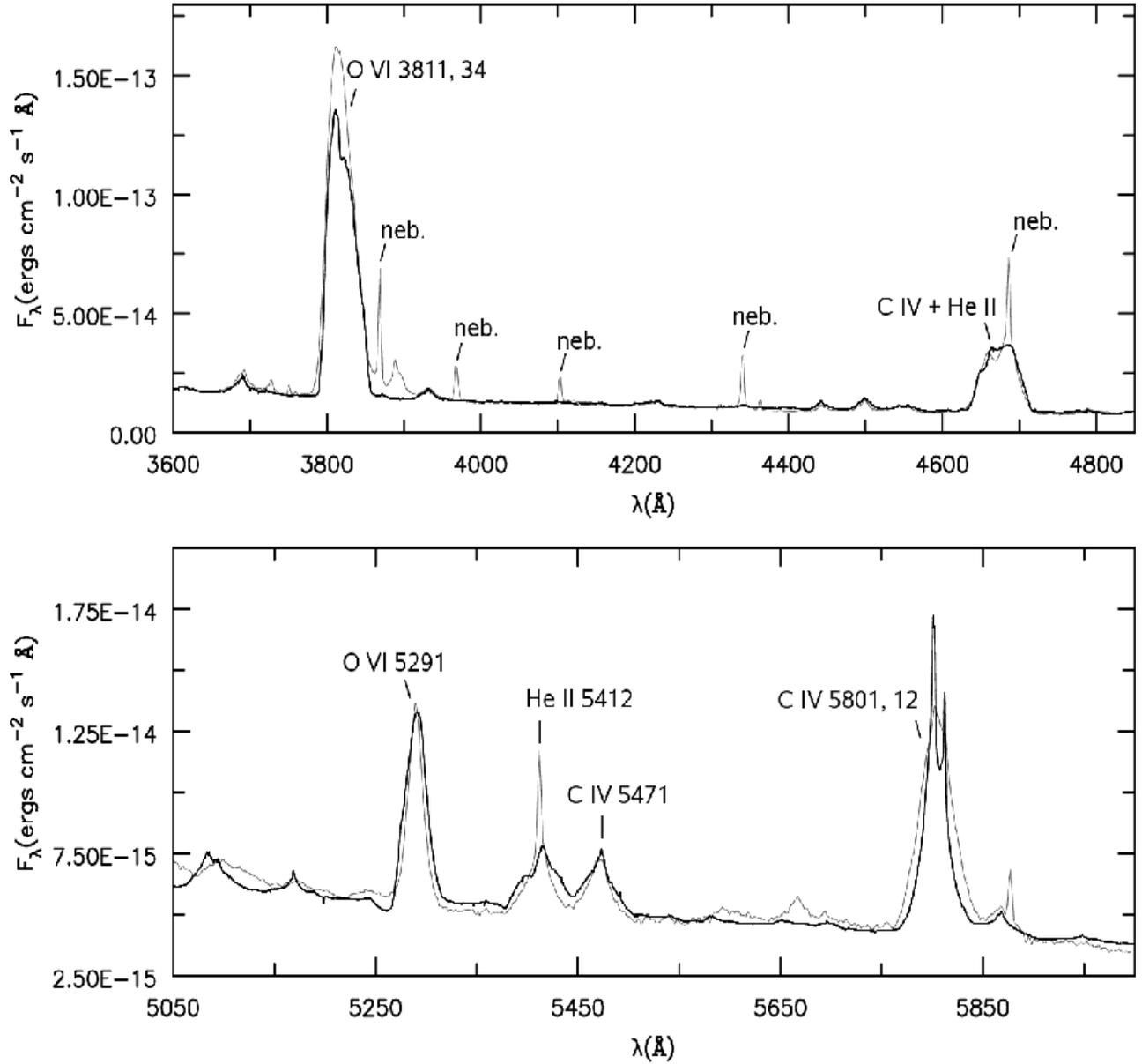}
\caption{NGC 6905 optical spectrum and our model (thick line).
Nebular lines are : [\ion{Ne}{3}] $\lambda 3868$, $\lambda 3969$, H$\delta$, H$\gamma$, 
\ion{He}{2} $\lambda 4686$ and \ion{He}{2} $\lambda 5412$ (contamination).}
\label{ngc6905_optical}
\end{figure} 

\begin{figure}[!ht] 
\centering  
\includegraphics[width=16cm]{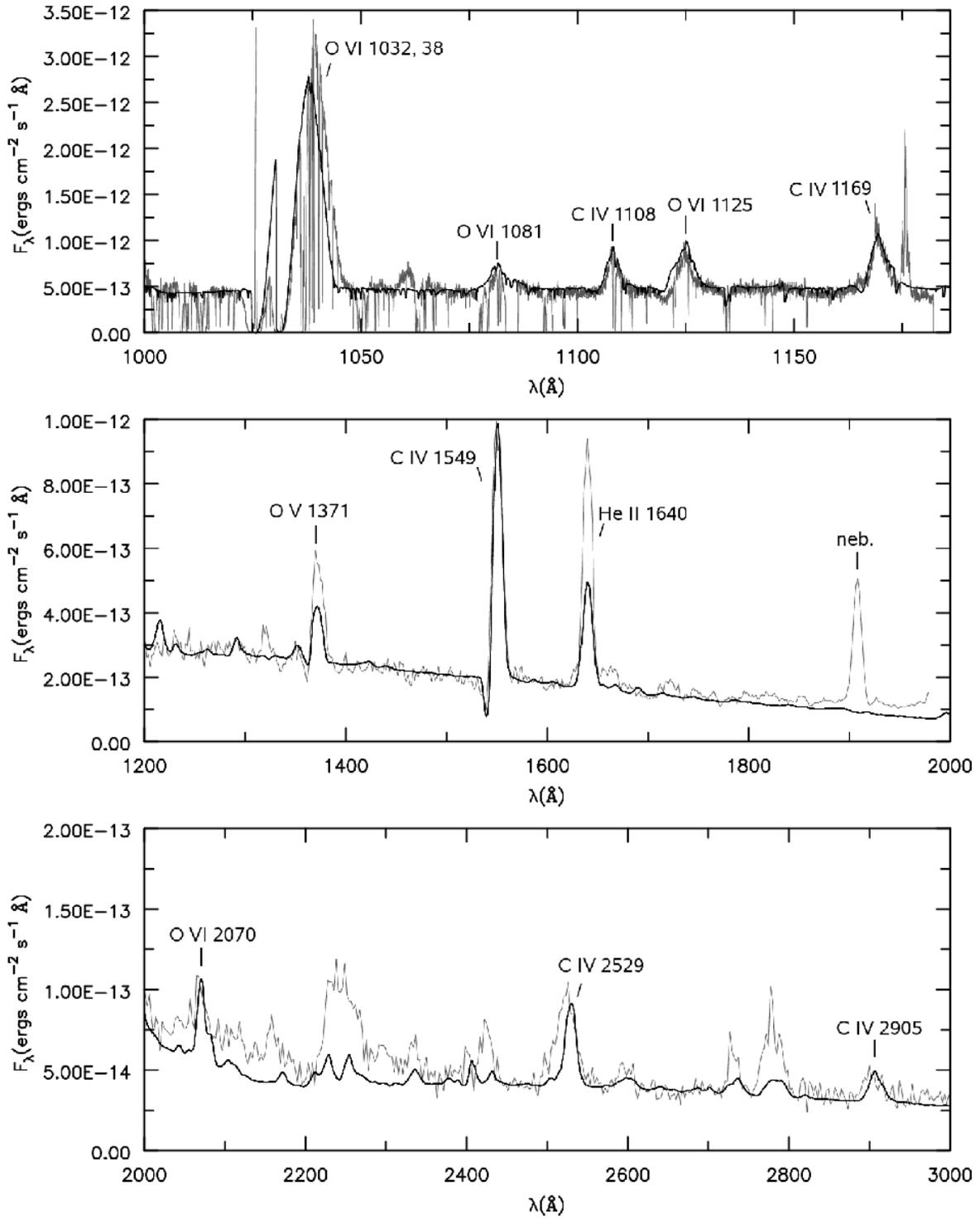}
\caption{NGC 6905 ultraviolet spectra and our model (thick line).
Top : FUSE spectrum. Other panels : Low resolution IUE spectra. }
\label{ngc6905_uv}
\end{figure} 

\begin{figure}[!ht] 
\centering  
\includegraphics[width=14cm]{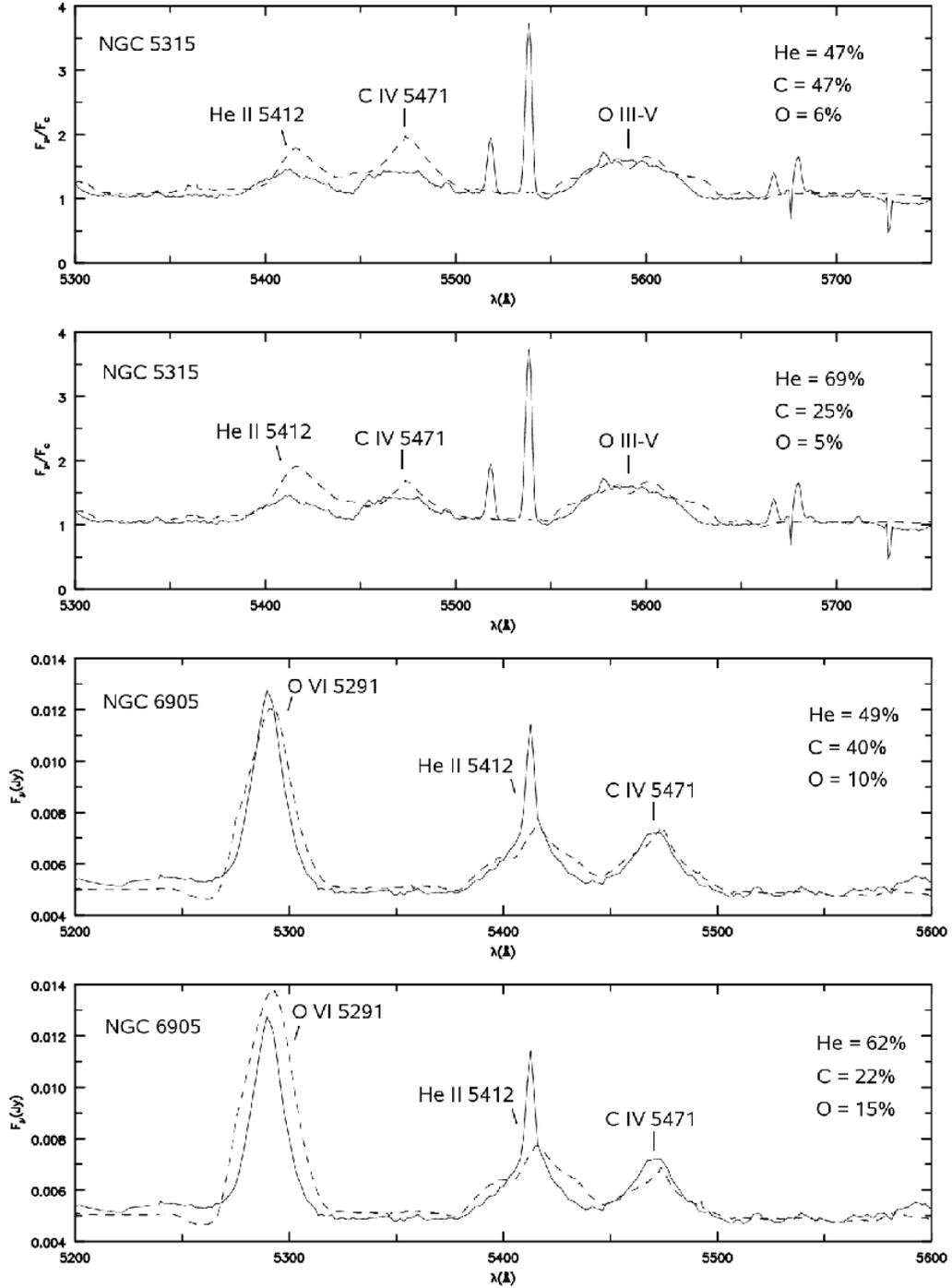}
\caption{Illustration of the effect of variations in the
$\beta_{C}/\beta_{He}$ ratio (mass fraction) on the theoretical spectrum 
of NGC 5315 and NGC 6905. Solid lines are the observational data and dashed lines are the
models. Top panel (NGC 5315) : $\beta_{C} = \beta_{He} = 47\%$ and $\beta_O = 6\%$. 
2nd panel (NGC 5315) : $\beta_{C} = 25\%$, $\beta_{He} = 69\%$ and $\beta_O = 5\%$. 
3rd panel (NGC 6905) : $\beta_{C} = 40\%$, $\beta_{He} = 49\%$ and $\beta_O = 10\%$. 
4th panel (NGC 6905) : $\beta_{C} = 22\%$, $\beta_{He} = 62\%$ and $\beta_O = 15\%$.}
\label{abund}
\end{figure} 

\clearpage
\begin{figure}[!ht] 
\centering  
\includegraphics[width=17cm, height=17cm]{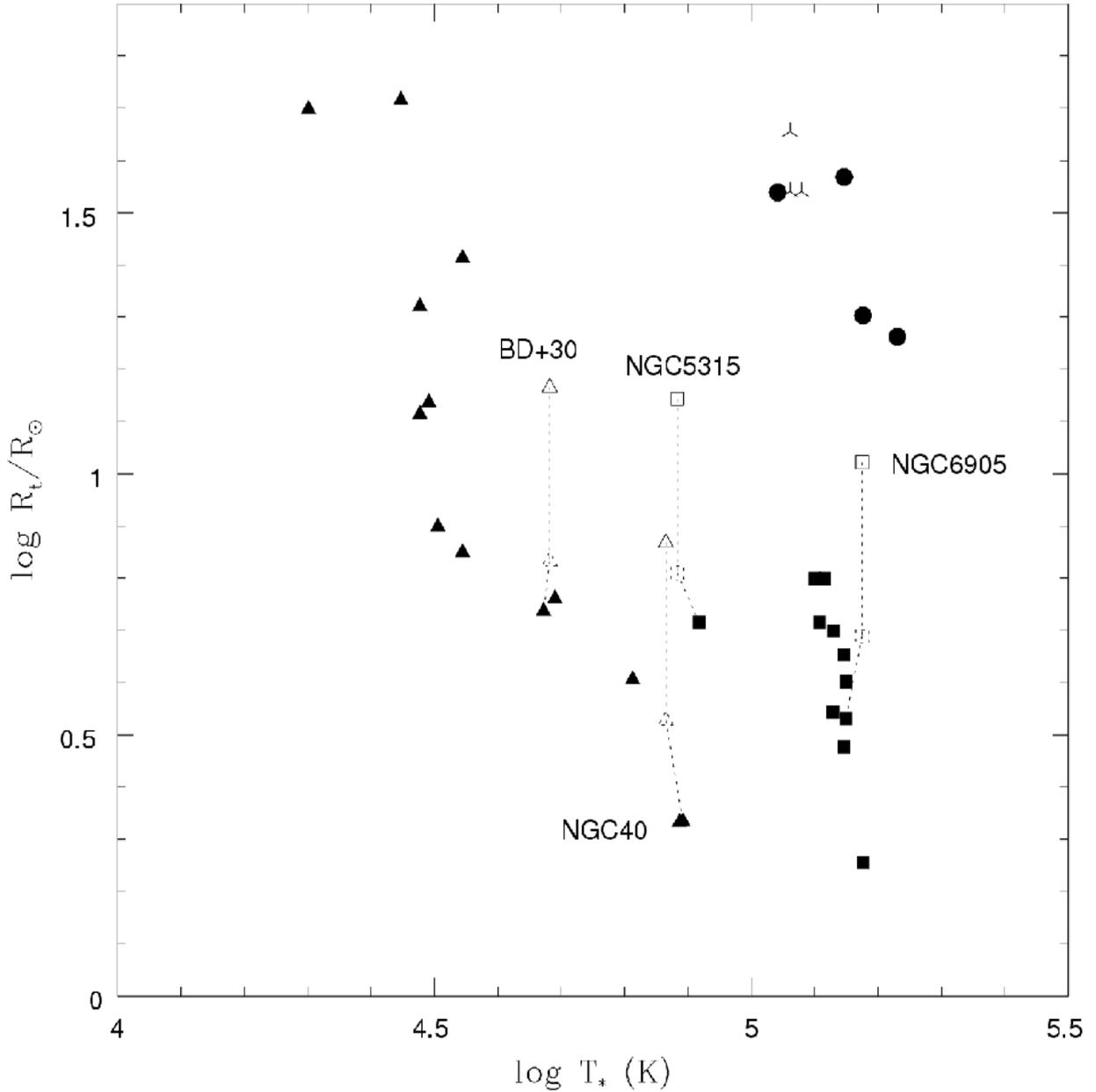}
\caption{Evolutionary sequence in the Log $R_{T} \times $ Log $T_{*}$
diagram. Literature data are represented by filled symbols (see Koesterke 2001 and references therein). 
[WCL] and early-type [WR] stars are represented by triangles and squares, respectively.
Crosses indicate [WELS] and PG 1159 stars are represented by circles. Open solid symbols 
represent our clumped models (with $f=0.1$) and dashed symbols our un-clumped models. 
Dashed lines connect them and previous results of the Potsdam group.}
\label{rt}
\end{figure} 

\begin{figure}[!ht] 
\centering  
\includegraphics[width=17cm, height=17cm]{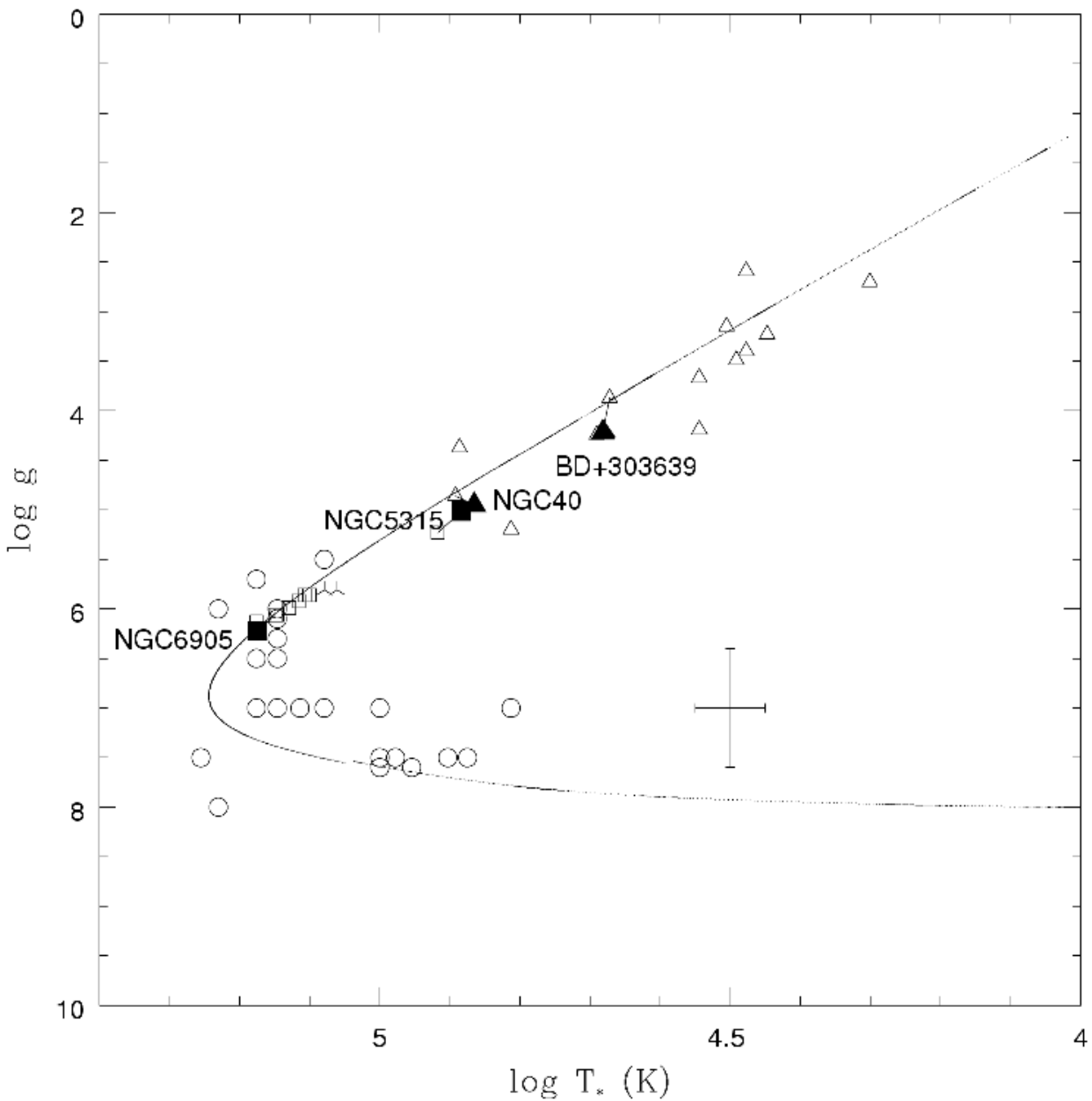}
\caption{Evolutionary sequence in the HR diagram (Log $g \times$ Log $T_{*}$) 
and values derived by means of non LTE atmosphere models. 
The track is from Althaus et al. (2005) for a 2.7$M_{\odot}$ main 
sequence star that evolved to a 0.5885$M_\odot$ remnant. [WCL] stars are
represented by open triangles, early-type [WR] stars by open squares, 
[WELS] by crosses and PG 1159 by open circles. Our data 
correspond to filled symbols. Vertical error bars represent an error 
in distance of a factor of $\pm$2, or analogously, a typical spectral uncertainty 
in the determination of Log $g$ (see H\"ugelmeyer et al. 2005). Horizontal 
error bars illustrate an uncertainty of 10\% in temperature determination.
Solid lines connect our results with previous works.}
\label{HR}
\end{figure} 

\end{document}